\newcommand{\Title}{Active inference, eye movements and oculomotor delays}%
\newcommand{\AuthorA}{Laurent U.~Perrinet}%
\newcommand{\AuthorB}{Rick A.~Adams} %
\newcommand{\AuthorC}{Karl J.~Friston} %
\newcommand{\Address}{Institut de Neurosciences de la Timone \\ CNRS / Aix-Marseille Universit\'e - Marseille, France}%
\newcommand{\AddressBis}{The Wellcome Trust Centre for Neuroimaging \\ University College London - London, UK.}%
\newcommand{\Website}{http://invibe.net/LaurentPerrinet}%
\newcommand{\Email}{Laurent.Perrinet@univ-amu.fr}%
\newcommand{\Abstract}{This paper considers the problem of sensorimotor delays in the optimal control of (smooth) eye movements under uncertainty. Specifically, we consider delays in the visuo-oculomotor loop and their implications for active inference. Active inference uses a generalisation of Kalman filtering to provide Bayes optimal estimates of hidden states and action in generalized coordinates of motion. Representing hidden states in generalized coordinates provides a simple way of compensating for both sensory and oculomotor delays. The efficacy of this scheme is illustrated using neuronal simulations of pursuit initiation responses, with and without compensation. We then consider an extension of the generative model to simulate smooth pursuit eye movements --- in which the visuo-oculomotor system believes both the target and its centre of gaze are attracted to a (hidden) point moving in the visual field. Finally, the generative model is equipped with a hierarchical structure, so that it can recognise and remember unseen (occluded) trajectories and emit anticipatory responses. These simulations speak to a straightforward and neurobiologically plausible solution to the generic problem of integrating information from different sources with different temporal delays and the particular difficulties encountered when a system --- like the oculomotor system --- tries to control its environment with delayed signals. }%
\newcommand{\Keywords}{Oculomotor delays; tracking eye movements; smooth pursuit eye movements; generalized coordinates; perception; Bayesian filtering; variational free energy; active inference}%
\newcommand{\Acknowledgments}{
LuP was supported by EC IP project FP6-015879, ``FACETS'' and FP7-269921, ``BrainScaleS'' and by the Wellcome Trust Centre for Neuroimaging. RAA and KJF are supported by the Wellcome Trust Centre for Neuroimaging. %
}%
\documentclass[a4paper]{article} 
\usepackage[english]{babel}%
\usepackage{amsmath,amsfonts,amsthm}			
\usepackage{tikz}
\usetikzlibrary{calc}
\usepackage{graphicx}					

\usepackage[natbib=true,
			style=authoryear,
			citestyle=authoryear-comp,
			sortcites=true,
			dashed=false,
			uniquename=false,
			block=space,
			backend=bibtex,
			doi=false,isbn=false,url=false,
			]{biblatex}%
\usepackage{siunitx}%
\newcommand{\Hz}{\si{\hertz}}%
\newcommand{\ms}{\si{\milli\second}}%
\newcommand{\meter}{\si{\meter}}%
\newcommand{\s}{\si{\second}}%
\usepackage{bm}
\usepackage{amsmath,amsfonts,amsthm}			
\usepackage[unicode,linkcolor=blue,citecolor=blue,filecolor=black,urlcolor=blue,pdfborder={0 0 0}]{hyperref}%
\hypersetup{%
pdftitle={\Title},%
pdfauthor={\AuthorA < \Email > \Address - \Website},%
pdfkeywords={\Keywords},%
pdfsubject={\Acknowledgments}%
}%
\usepackage{authblk}%
\title{\Title}%
\author[1,2]{\AuthorA }
\author[2]{\AuthorB } 
\author[2]{\AuthorC } 
\affil[1]{\Address}
\affil[2]{\AddressBis}

\date{}
\begin{document}%
\maketitle%
\section*{Keywords}%
\Keywords
\section*{Abstract}
\Abstract %
\section{Introduction}%
\subsection{Problem statement}%
This paper considers optimal motor control and the particular problems caused by the inevitable delay between the emission of motor commands and their sensory consequences. This is a generic problem that we illustrate within the context of oculomotor control where it is particularly prescient (see for instance~\citep{Nijhawan08} for a review). Although we focus on oculomotor control, the more general contribution of this work is to treat motor control as a pure inference problem. This allows us to use standard (Bayesian filtering) schemes to resolve the problem of sensorimotor delays --- by absorbing them into a generative or forward model. Furthermore, this principled and generic solution has some degree of biological plausibility because the resulting active (Bayesian) filtering is formally identical to predictive coding, which has become an established metaphor for neuronal message passing in the brain. We will use oculomotor control as a vehicle to illustrate the basic idea using a series of generative models of eye movements --- that address increasingly complicated aspects of oculomotor control. In short, we offer a general solution to the problem of sensorimotor delays in motor control --- using established models of message passing in the brain --- and demonstrate the implications of this solution in the particular setting of oculomotor control.

\begin{figure}
\begin{center}
   \includegraphics[width=\columnwidth]{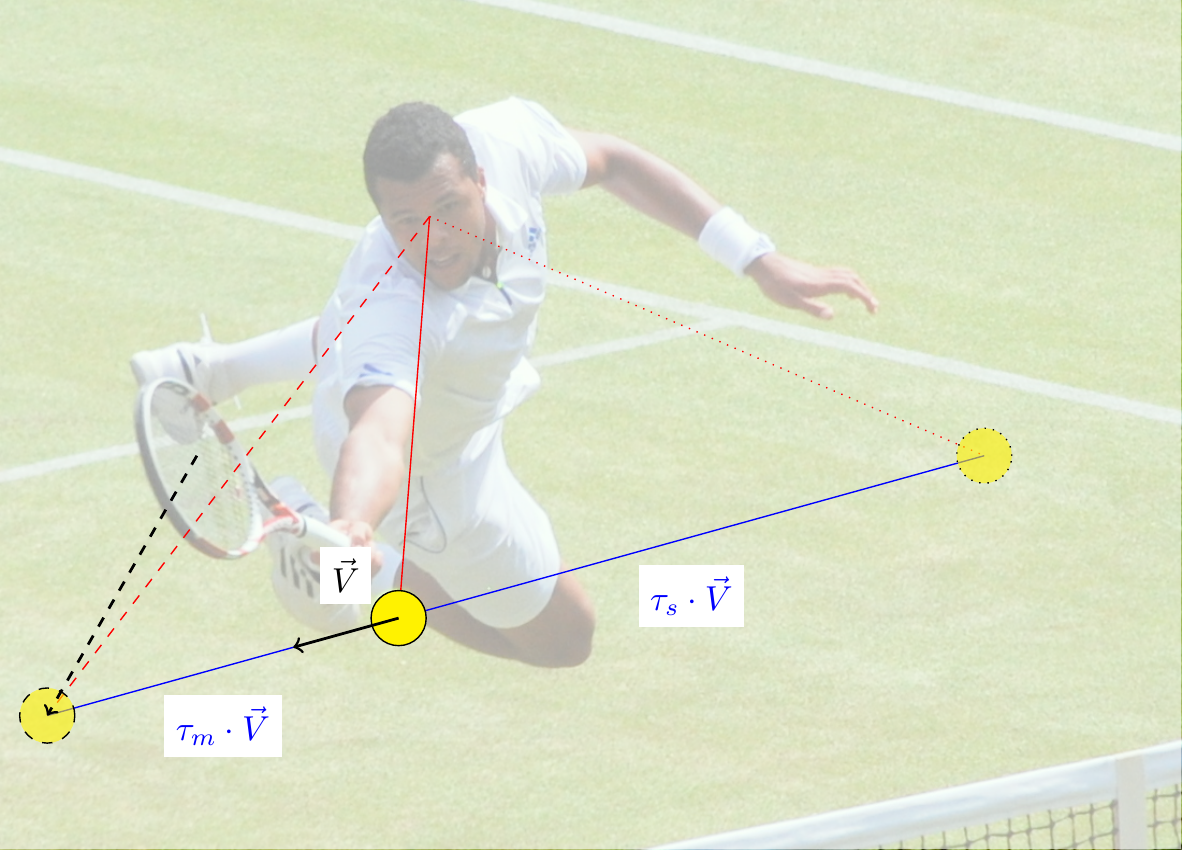}%
\end{center}%
\caption{%
Problem statement: optimal motor control under axonal delays. The central
nervous system has to contend with axonal delays, both at the sensory and the
motor levels. For instance, in the human visuo-oculomotor system, it takes
approximately $\tau_s=50~\ms$ for the retinal image to reach the visual areas
implicated in motion detection, and a further $\tau_m=40~\ms $ to reach the
oculomotor muscles. As a consequence, for a tennis player trying to intercept a
ball at a speed of $20~\meter\cdot \s^{-1}$, the sensed physical position is $1~\meter$
behind the true position (as represented here by $\tau_s \cdot \vec{V}$), while
the position at the moment of emitting the motor command will be $.8~\meter$ ahead
of its execution ($\tau_m \cdot \vec{V}$). Note that while the actual position
of the ball when its image produced by the photoreceptors on the retina hits 
visual areas is approximately at
$45$ degrees of eccentricity (red dotted line), the player's gaze is directed to
the ball at its \emph{present} position (red line), in anticipatory fashion.
Optimal control directs action (future motion of the eye) to the expected
position (red dashed line) of the ball in the future --- and the racket (black
dashed line) to the expected position of the ball when motor commands reach the
periphery (muscles).}%
\label{fig:figure0}
\end{figure}

The oculomotor system produces eye movements to deploy sensory (retinal)
epithelia at very fast timescales. In particular, changes in the
position of a visual object are compensated for with robust and rapid
eye movements, such that the object is perceived as invariant, despite
its motion~\citep{Ilg97,Lisberger87}. This near-optimal control
is remarkable, given the absence of any external clock to coordinate dynamics in
different parts of the visual-oculomotor system. An important constraint, in
this setting, is axonal conduction, which produces delays in sensory and motor
signalling within the oculomotor system. 
Figure~\ref{fig:figure0} shows that in humans, for example, retinal signals
arriving at motion processing areas report the state of affairs at least about
50~\ms\ ago, while the action that follows is executed at least 40~\ms\ in the
future~\citep{Inui06}; for a review, see~\citep{Masson10a}.  Different sources of delays exist -- such as the
biomechanical delay between neuromuscular excitation and eye movement. Due to
these delays, the human smooth pursuit system responds to unpredictable stimuli
with a minimum latency of around 100~\ms~\citep{Wyatt87}. In addition, these
delays may produce oscillations about a constant velocity
stimulus~\citep{Robinson86,Robinson65}, whose amplitude and frequency can be
altered by artificially manipulating the feedback ~\citep{Goldreich92}.

Eye movements can anticipate predictable stimuli, such as the sinusoidal
movement of a pendulum~\citep{Barnes91,Dodge30,Westheimer54}; for a review,
see~\citep{Barnes08}. Interestingly, ocular tracking can compensate for
sensorimotor delays after around one or two periods of sinusoidal motion --
producing a tracking movement with little discernible delay~\citep{Barnes91}.
This suggests that the oculomotor system can use sensory information from the
past to predict its future sensory states (including its actions), despite the
fact that these sensory changes can be due to both movement of the stimulus and
movement of the eyes. The time taken to compensate for delays increases with the
unpredictability of the stimulus~\citep{Michael66}, though the system can adapt
quickly to complex waveforms, with changes of velocity~\citep{Barnes02}, single
cycles~\citep{Barnes00} or perturbed periodic waves -- where subjects appear to
estimate their frequency using an average over recent cycles~\citep{Collins09}.
Further studies suggest that different sources of information, such as auditory
or verbal cues~\citep{Kowler89} or prior knowledge about the nature of sensory
inputs~\citep{Montagnini06} can evoke anticipatory eye movements.

The aim of this work was to establish a principled model of optimal visual motion processing and oculomotor control in the context of sensorimotor delays. Delays are often ignored in treatments of the visual-oculomotor system; however, they are crucial to understanding the system's dynamics. For instance, delays may be important for understanding the pathophysiology of impaired oculomotor control: schizophrenic smooth pursuit abnormalities are due to impairments of the predictive (extra-retinal) motion signals that are required to compensate for sensorimotor delays~\citep{Nkam10,Thaker99}. Surprisingly, delays may also explain a whole body of visual illusions~\citep{Changizi2001Perceiving,Changizi02,Changizi08,Vaughn13}, even for visual illusions that involve a static display. Delays are also an important consideration in control theory and engineering. Finally, neuronal solutions to the delay problem speak to the representation of time in the brain, which is essential for the proper fusion of information in the central nervous system.

\subsection{Existing solutions and the proposed hypothesis}

A principled approach to optimal oculomotor control is provided by
Bayesian filtering schemes that use probabilistic representations to
estimate visual and oculomotor states. These states are \emph{hidden};
i.e., they cannot be measured directly. A popular scheme for linear
control problems is the Kalman filter~\citep{Kalman60}. The Kalman scheme
can be extended to accommodate biomechanical constraints, such as
transmission delays (e.g., fixed-lag smoothers). However, their
solutions can become computationally complex when delays are large in
relation to discretisation time and are not biologically plausible. We have previously considered
generalized Bayesian filtering in continuous time as a metaphor for
action and perception. 
This approach has been applied to eye movements~\citep{Friston10a}
and saccades in particular~\citep{Friston12}. However, these
applications ignored sensorimotor delays and their potentially
confounding effects on optimal control.

Crucially, the active inference schemes we have considered
previously are formulated using representations in \emph{generalized
coordinates of motion}; that is, states (such as position) are
represented along with their higher order temporal derivatives (such as
speed, acceleration, jerk, etc). This means that one has an implicit
representation of hidden states in the recent past and future that can
be used to finesse the problems of delays. For example, it has been
shown that acceleration is a necessary component of the predictive drive
to eye movements~\citep{Bennett07}. In brief, generalized
representations can be projected to the past and to the future using
simple (linear) mixtures of generalized motion. Note that a
representation of generalized motion can be explicit or implicit by
using a population coding scheme -- as has been demonstrated for
acceleration~\citep{Lisberger99}. Representations of
generalized motion may be important for modelling delays when
integrating information in the brain from distal sources -- such as
other cortical columns in the same cortical area or other areas that are
connected with fixed but different delays~\citep{Roelfsema97}. The integration of
information over time becomes particularly acute in motor control, where
the products of sensory processing couple back to the sampling of
sensory information through action.

In the context of action, acted inference finesses the problems with
delayed control signals in classical formulations of motor control by
replacing command signals with descending cortico-spinal predictions. 
For instance, the location of receptive fields in the parietal cortex in 
monkeys is shown to shift transiently before an eye movement~\citep{Duhamel92}. %
These predictions are fulfilled at the peripheral level, using fast
closed loop mechanisms (peripheral reflex arcs). In principle, ``these
predictions can anticipate delays if they are part of the generative
model,''~\citep{Friston11}: However, this anticipation has never been
demonstrated formally. Here, we show how generalized Bayesian filtering
-- as used in active inference -- can compensate for both sensory and
motor delays in the visual-oculomotor loop.

It is important to mention what this work does not address. First, we
focus on tracking eye movements (pursuit of a single dot stimulus for a
monocular observer with fixed head position): we do not consider other
types of eye movements (vergence, saccades, or the vestibulo-ocular
reflex). Second, we take an approach that complements existing models,
such as those of~\citep{Robinson86} and~\citep{Krauzlis89}. Existing 
models account for neurophysiological and behavioural
data by refining block-diagram models of oculomotor control to describe
\emph{how} the system might work. We take a more generic approach, in
which we define the imperatives for any system sampling sensory data,
derive an optimal oculomotor control solution and show \emph{why} this
solution explains the data. Although the two approaches should be
consistent, ours offers a principled approach to identifying the
necessary solutions (such as predictive coding) to a given problem
(oculomotor delays). We hope to demonstrate the approach by modelling
pursuit initiation and smooth pursuit -- and then consider the
outstanding issue of anticipatory responses: in previous treatments~\citep{Robinson86}, 
``{[}anticipation{]} has not been adequately modelled and no such attempt 
is offered (\ldots) only unpredictable movements are considered.''

\subsection{Outline}
The main contributions of our work are described in the subsequent five sections. First, section~\ref{sec:active-inference} summarises the basic theory behind active inference and attempts to link generalized filtering to conventional Bayesian filters used in optimal control theory. This section then considers neurobiological implementations of generalized filtering, in terms of predictive coding in generalized coordinates of motion. This formulation allows us to consider the problem of delayed sensory input and motor output in section~\ref{sec:delay} -- and how this problem can be finessed in a relatively straightforward way using generalized representations. %
Having established the formal framework
(and putative neuronal implementation), the final three sections deal with
successively harder problems in oculomotor control. We start in
Section~\ref{sec:pursuit} by considering \emph{pursuit initiation} using a
simple generative model of oculomotor trajectories. Using simulations, we
consider the impact of motor delays, sensory delays and their interaction on
responses to a single sweep of a visual target. The subsequent section turns to
\emph{smooth pursuit eye movements} -- using a more sophisticated generative
model of oculomotor trajectories, in which prior beliefs about eye movements
enable the centre of gaze to predict target motion using a virtual or fictive
target (see Section~\ref{sec:spem}). In the final section, we turn to
hierarchical models of target trajectories that have explicit memories of hidden
dynamics, which enable anticipatory responses (see Section~\ref{sec:anticip}).
These responses are illustrated using simulations of anticipatory pursuit
movements using (rectified) hemi-sinusoidal motion. In short, these theoretical
considerations lead to a partition of stimulus--bound eye movements into pursuit
initiation, smooth pursuit and anticipatory pursuit, where each mode of
oculomotor control calls on formal additions to the underlying generative model;
however, they all use exactly the same scheme and basic principles. Where
possible, we try to simulate classic empirical results in this field -- at least
heuristically.

In short, these theoretical considerations lead to a partition of stimulus-bound
eye movements into pursuit initiation, smooth pursuit and anticipatory pursuit,
where each mode of oculomotor control calls on formal additions to the
underlying generative model. However, these models all use exactly the same
scheme and basic principles; in particular, they all use the same solution to
the oculomotor delay problem. These simulations illustrate that the active
inference scheme can reproduce classical empirical results in three distinct
experimental contexts.

\begin{figure}
 \centerline{%
\includegraphics[width=\columnwidth]{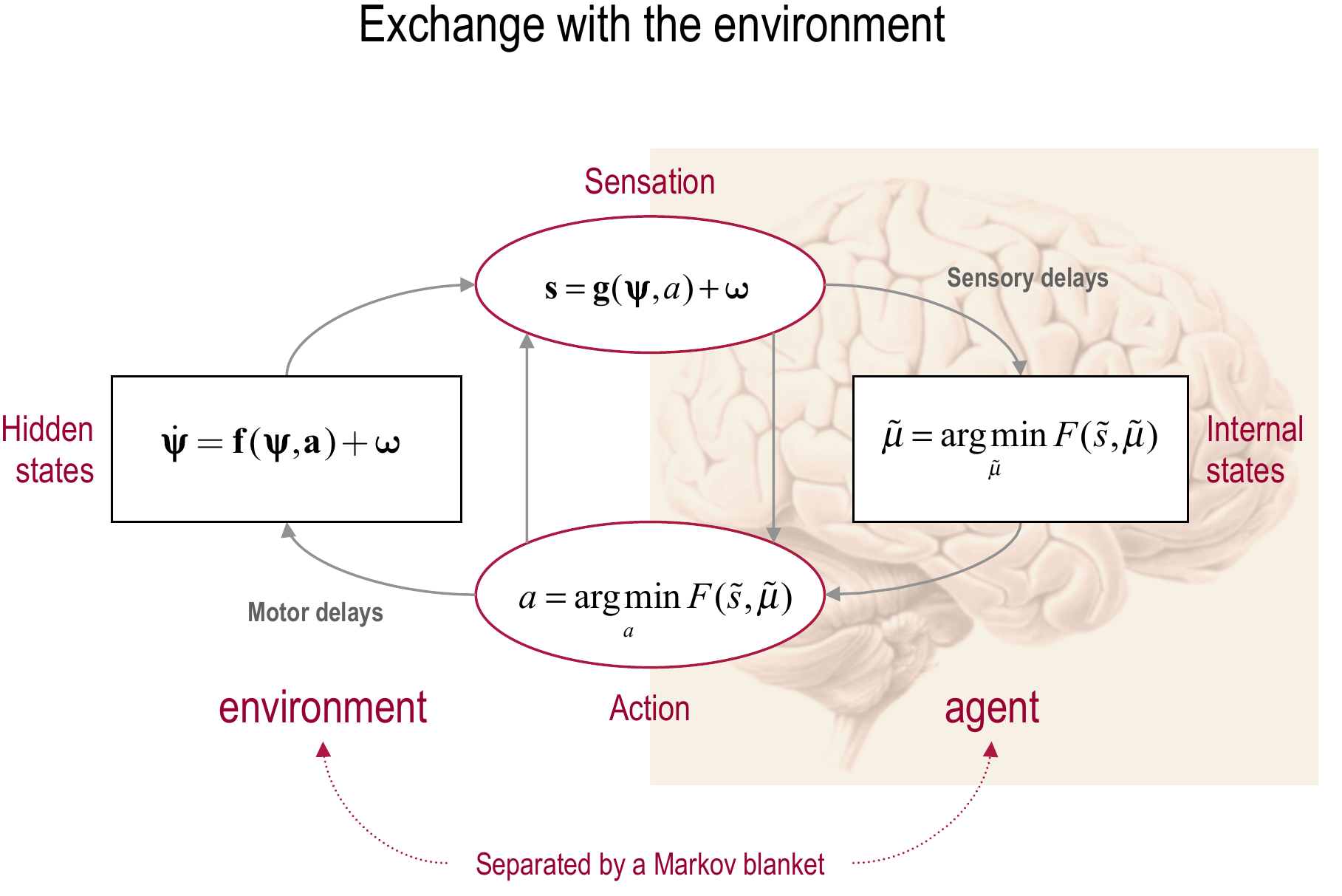}%
}%
\caption{This schematic shows the dependencies among various
quantities modelling exchanges of an agent with the environment. It
shows the states of the environment and the system in terms of a
probabilistic dependency graph, where connections denote directed
dependencies. The quantities are described within the nodes of this
graph -- with exemplar forms for their dependencies on other variables
(see main text). Hidden (external) and internal states of the agent are
separated by action and sensory states. Both action and internal states
-- encoding a conditional probability density function over hidden
states -- minimise free energy. Note that hidden states in the real
world and the form of their dynamics can be different from that assumed
by the generative model; this is why hidden states are in bold. See main
text for further details.
}%
\label{fig:figure1}
\end{figure}

\section{From predictive coding to active inference}
\label{sec:active-inference}
This section sets out the basic theory, before applying it to the
special problem of oculomotor delays in the following sections. We first
introduce the general framework of active inference in terms of
generalized Bayesian filtering and variational free energy minimisation.
In brief, active inference can be regarded as equipping standard
Bayesian filtering schemes with classical reflex arcs that enable action, such as an eye movement, %
to fulfil predictions about hidden states of the world. Second, we will
briefly describe the formalism of active inference in terms of
differential equations describing the dynamics of the world and internal
states of the visual-oculomotor system. The neurobiological
implementation of these differential equations is considered in terms of
predictive coding, which uses prediction errors on the motion of hidden
states -- such as the location of a visual target. In the next section,
we will turn to the special problem of oculomotor delays and how this
problem can be finessed using active inference in generalized
coordinates of motion. This solution will be illustrated in subsequent
sections using simulations of pursuit initiation responses and smooth
pursuit. Finally, we shall exploit the richness of hierarchical
generative models -- which underlie active inference -- to illustrate
anticipatory eye movements that cannot be explained by simply
compensating for oculomotor delays.

\subsection{From free energy to generalized filtering}

The scheme used here to model oculomotor behaviour has been used to
model several other processes and paradigms in neuroscience. This active inference scheme is based upon three assumptions: 

\begin{itemize}
\item
  The brain minimises the free energy of sensory inputs defined by a
  generative model.
\item
  The generative model used by the brain is hierarchical, nonlinear and
  dynamic.
\item
  Neuronal firing rates encode the expected state of the world, under
  this model.
\end{itemize}

The first assumption is the free energy principle, which leads to active
inference in the context of an embodied interaction of the system with
its environment; where the system can act to change its sensory inputs.
The free energy here is a variational free energy that provides a
computationally tractable upper bound on the negative logarithm of
Bayesian model evidence (see Appendix 1). In Bayesian terms, this means
that the brain maximises the evidence for its model of sensory inputs~\citep{Ballard83,Bialek01,Dayan95,Gregory80,Grossberg97,Knill04,Olshausen96}. This is the Bayesian brain hypothesis~\citep{Yuille06}. %
If we also allow action to maximise model evidence, we get active
inference~\citep{Friston10a}. Crucially, unlike conventional optimal
control schemes, there is no \emph{ad hoc} value or loss function guiding
action: Action minimises the free energy of the system's model. This
permits the application of standard Bayesian solutions and simplifies
the implicit neuronal architecture; for example, there is no need for an
efference copy signal~\citep{Friston11}. In this setting, desired movements
are specified in terms of prior beliefs about state transitions or the
motion of hidden states in the generative model. Action then realises
prior beliefs (policies) by sampling sensory data that provides evidence
for those beliefs. %

The second assumption above is motivated by noting that the world is
both dynamic and nonlinear -- and that hierarchical structure emerges
inevitably from a separation of temporal scales~\citep{Ginzburg55,Haken83}. The third assumption is the Laplace assumption that, in terms of
neural codes, leads to the \emph{Laplace code}, which is arguably the
simplest and most flexible of all neural codes~\citep{Friston09a}.  %
In brief, the Laplace code means that probabilistic representations are encoded explicitly by synaptic activity in terms of their mean or expectation (while the second order statistics such as dispersion or precision are encoded implicitly in terms of synaptic activity and efficacy). This limits the representation of hidden states to continuous variables, as opposed to discrete states; however, this is appropriate for most aspects of sensorimotor processing. Furthermore, it finesses the combinatoric explosion associated with discrete state space models. Restricting probabilistic representations to a Gaussian form clearly precludes multimodal representations. Having said this, the hierarchical form of the generative models allows for fairly graceful modelling of nonlinear effects (such as shadows and occlusions). For example, a Gaussian variable at one level of the model may enter the lower levels in highly non-linear way --- we will see examples of this later. %
See Appendix 2 for a motivation of the Laplace assumption from basic principles.

Under these assumptions, action and perception can be regarded as the
solutions to coupled differential equations describing the dynamics of
the real world (the first pair of equations) and the behaviour of an
agent (the second pair of equations); expressed in terms of action and
internal states that encode conditional expectations about hidden states
of the world~\citep{Friston10a}:

\begin{align}
\bm{s} &= \bm{g(x, \nu, a) + \omega_\nu} \nonumber \\%
\bm{\dot{x}} &= \bm{f(x, \nu, a) + \omega_x}  \nonumber  \\%
  \label{eq:1} \\%
\dot{a} &= -\partial_a F(\tilde{s}, \tilde{\mu}) \nonumber   \\%
\dot{\tilde{\mu}} &= \mathcal{D} \tilde{\mu} - \partial_{\tilde{\mu}} F(\tilde{s}, \tilde{\mu})  \nonumber  %
\end{align}%

For clarity, real-world states are written in boldface, while internal states of the agent are in italics: Here $\bm{(s, x, \nu, a)}$ denote sensory input, hidden states, hidden causes and action in the real world, respectively. The variables in the second pair of equations $(\tilde{s}, \tilde{\mu}, a)$ correspond to generalized sensory input, conditional expectations and action. %
Generalized coordinates of motion, denoted by the \textasciitilde{} notation, correspond to a vector representing the different orders of motion of a variable: position, velocity, acceleration, and so on~\citep{Friston10d}. Using the Lagrangian notation for temporal derivatives, we get e.~g. for $s$: $\tilde{s} = (s, s^{\prime}, s^{\prime\prime}, \ldots)$. %
In the absence of delays $\tilde{s}(t) = \bm{\tilde{s}}(t)$ the agent receives instantaneous sensations from the real world. The differential equations above are coupled because sensory states depend upon action through hidden states and causes $\bm{(x, \nu)}$ while action $a(t) = \bm{a}(t)$ depends upon sensory states through internal states $\tilde{\mu}$.

By explicitly separating real-world states --hidden from the agent-- to its internal states,  one can clearly separate the generative model from the updating scheme that allows to minimise the agent's free-energy: The first pair of coupled stochastic differential equations describes the dynamics of hidden states and causes in the world and how they generate sensory states. These equations are stochastic because sensory states and the motion of hidden states are subject to random fluctuations $(\bm{\omega_x, \omega_\nu})$. 

The second pair of differential equations corresponds to action and perception respectively -- they constitute a (generalized) gradient descent on variational free energy. The differential equation describing changes in conditional expectations (perception) is known as \emph{generalized filtering} or predictive coding and has the same form as standard Bayesian (Kalman-Bucy) filters -- see also~\citep{Beal03,Rao99}. %
The first term is a prediction based upon a differential operator $\mathcal{D}$ that returns the generalized motion of the conditional expectations; namely the vector of velocity, acceleration, jerk and so on -- such that  $\mathcal{D}\tilde{\mu} = (\mu^{\prime}, \mu^{\prime\prime}, \mu^{\prime\prime\prime}, \ldots)$. However, the expected velocity is not the velocity of the expectation and comprises both prediction and update terms: The second term reflects this correction and ensures the changes in conditional expectations are Bayes-optimal predictions of hidden states of the world -- in the sense that they maximise the free energy bound on Bayesian model evidence. See Figure~\ref{fig:figure1} for a schematic summary of the implicit conditional dependencies implied by Equation~\ref{eq:1}.

\subsection{Hierarchical form of the generative model}

\begin{figure}
\centerline{%
\includegraphics[width=\columnwidth, clip, trim = 2.3cm 2cm 1.1cm 1.2cm, page=2]{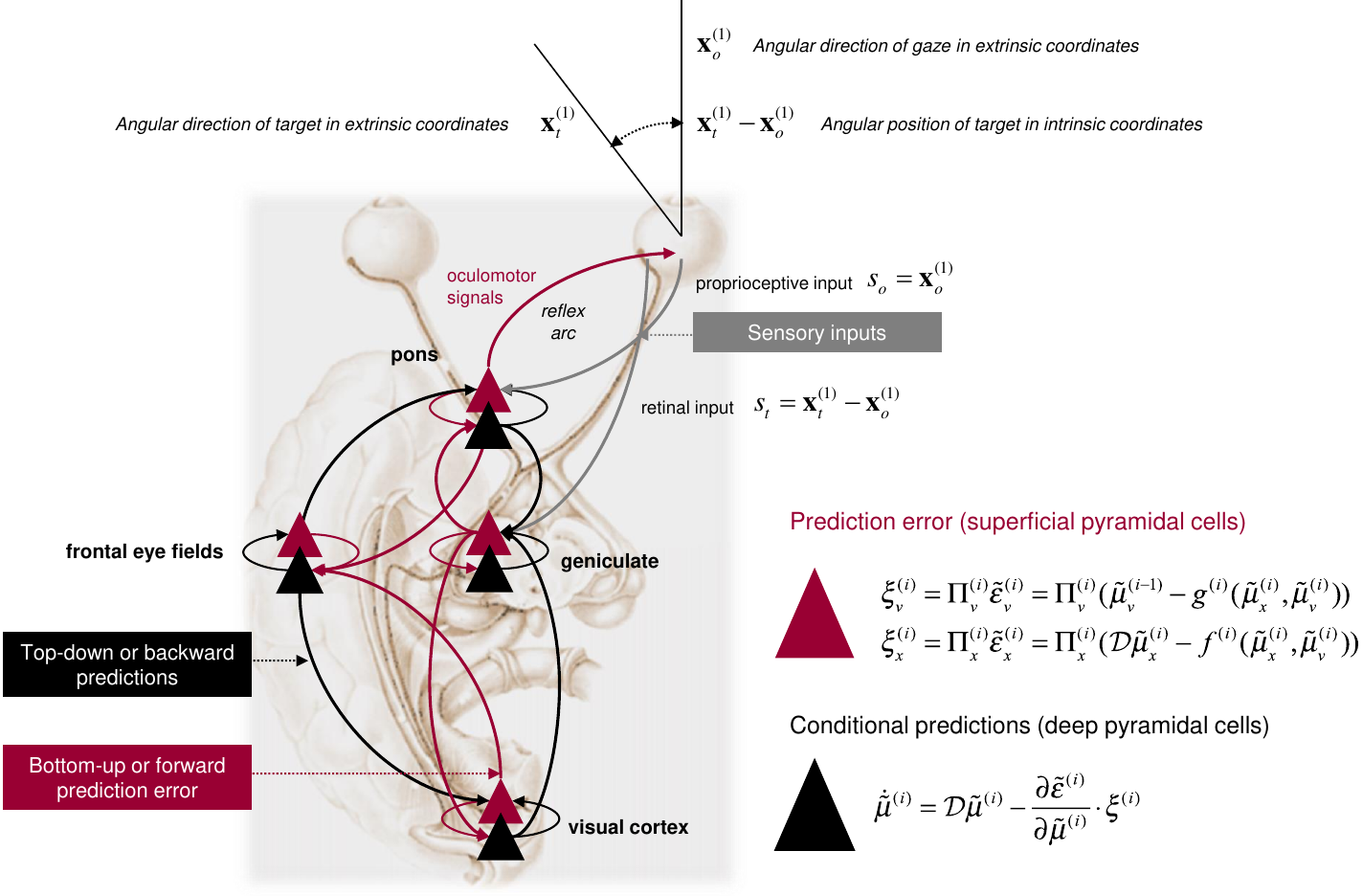}  
}%
\caption{Schematic detailing a neuronal message passing scheme
(generalized Bayesian filtering or predictive coding) that optimises
conditional expectations about hidden states of the world, given sensory
(visual) data and the active (oculomotor) sampling of those data. This
diagram shows the speculative cells of origin of forward driving
connections (in red) that convey prediction error from a lower area to a
higher area and the backward connections (in black) that construct
predictions~\citep{Mumford92}. These predictions try to explain away
prediction error in lower levels. In this scheme, the sources of forward
and backward connections are superficial (red) and deep (black)
pyramidal cells respectively. The equations on the right represent a
generalized descent on free energy under the hierarchical model
described in the main text -- this can be regarded as a generalisation
of predictive coding or Kalman filtering: see~\citep{Friston08b}.
State-units are in black and error-units are in red. Here, we have
placed different levels of some hierarchical model within the
visual-oculomotor system. Visual input arrives in an intrinsic (retinal)
frame of reference that depends upon the angular position of a stimulus
and the direction of gaze. Exteroceptive input is then passed to the
lateral geniculate nuclei (LGN) and to higher visual and prefrontal
(e.g., motion sensitive, such as the frontal eye field) areas in the
form of prediction errors. Crucially, proprioceptive sensations are also
predicted, creating prediction errors at the level of the cranial nerve
nuclei (pons). The special aspect of these proprioceptive prediction
errors is that they can be resolved through classical reflex arcs -- in
other words, they can elicit action to change the direction of gaze and
close the visual--oculomotor loop.}%
\label{fig:figure2}
\end{figure}

To perform simulations using this scheme, one simply integrates or
solves Equation~\ref{eq:1} to simulate (neuronal) dynamics that encode
conditional expectations and ensuing action. Conditional expectations
depend upon a generative model, which we assume has the following
(hierarchical) form

\begin{align}
s &= g^{(1)}(x^{(1)}, v^{(1)}) + \omega^{(1)}_\nu \nonumber \\%
\dot{x}^{(1)} &= f^{(1)}(x^{(1)}, v^{(1)}) + \omega_x^{(1)}  \nonumber  \\%
\vdots  \label{eq:2} \\%
\nu^{(i-1)} &= g^{(i)}(x^{(i)}, v^{(i)}) + \omega^{(i)}_\nu \nonumber \\%
\dot{x}^{(i)} &= f^{(i)}(x^{(i)}, v^{(i)}) + \omega_x^{(i)}  \nonumber  \\%
\vdots \nonumber  %
\end{align}%

Where $(i)$ indexes the level in the hierarchical model. Note that we denote the sensory layer as $i=0$, but this indexing is somewhat arbitrary. %
This equation is just a way of writing down a generative model that
specifies a probability density function over sensory inputs and hidden
states and causes. This probability density is needed to define the free
energy of sensory input: it is specified in terms of functions $(f^{(i)} , g^{(i)})$ and Gaussian assumptions about random fluctuations $(\omega^{(i)}_x, \omega^{(i)}_\nu)$ on
the motion of hidden states and causes. It is these that make the model
probabilistic -- they play the role of sensory noise at the first level
and induce uncertainty about states at higher levels. The precisions of
these fluctuations are quantified by $(\Pi^{(i)}_x, \Pi^{(i)}_\nu)$ 
which are defined as the inverse of the respective covariance matrices.

The deterministic part of the model is specified by nonlinear functions of hidden states and causes $(f^{(i)} , g^{(i)})$ that generate dynamics and sensory
consequences. Hidden causes link hierarchical levels, whereas hidden
states link dynamics over time. Hidden states and causes are abstract
quantities that the brain uses to explain or predict sensations -- like
the motion of an object in the field of view. In hierarchical models of
this sort, the output of one level acts as an input to the next. This
input can produce complicated convolutions with deep (hierarchical)
structure. We will see examples of this later in particular in the
context of anticipatory movements.

\subsection{Perception and predictive coding}

Given the form of the generative model (Equation~\ref{eq:2}) one can write down
the differential equations (Equation~\ref{eq:1}) describing neuronal dynamics in
terms of prediction errors on the hidden causes and states. These errors
represent the difference between conditional expectations and predicted
values, under the generative model (using $A \cdot B := A^T B$ for the scalar product and omitting higher-order terms):

\begin{align}
\dot{\tilde{\mu}}^{(i)}_x &= \mathcal{D} \tilde{\mu}^{(i)}_x + \frac{\partial \tilde{g}^{(i)}}{\partial \tilde{\mu}^{(i)}_x} \cdot \Pi^{(i)}_\nu \tilde{\varepsilon}^{(i)}_\nu + \frac{\partial \tilde{f}^{(i)}}{\partial \tilde{\mu}^{(i)}_x} \cdot \Pi^{(i)}_x \tilde{\varepsilon}^{(i)}_x -  \mathcal{D}  \Pi^{(i)}_x \tilde{\varepsilon}^{(i)}_x \nonumber \\%
\dot{\tilde{\mu}}^{(i)}_\nu &= \mathcal{D} \tilde{\mu}^{(i)}_\nu + \frac{\partial \tilde{g}^{(i)}}{\partial \tilde{\mu}^{(i)}_\nu} \cdot \Pi^{(i)}_\nu \tilde{\varepsilon}^{(i)}_\nu + \frac{\partial \tilde{f}^{(i)}}{\partial \tilde{\mu}^{(i)}_\nu} \cdot \Pi^{(i)}_x \tilde{\varepsilon}^{(i)}_x - \Pi^{(i+1)}_\nu \tilde{\varepsilon}^{(i+1)}_\nu \nonumber \\%
 \label{eq:3} \\%
\tilde{\varepsilon}^{(i)}_x &= \mathcal{D} \tilde{\mu}^{(i)}_x - \tilde{f}^{(i)}(\tilde{\mu}^{(i)}_x, \tilde{\mu}^{(i)}_\nu) \nonumber \\%
\tilde{\varepsilon}^{(i)}_\nu &= \tilde{\mu}^{(i-1)}_\nu - \tilde{g}^{(i)}(\tilde{\mu}^{(i)}_x, \tilde{\mu}^{(i)}_\nu)  \nonumber  %
\end{align}%

The quantities $\tilde{\varepsilon}^{(i)}$ correspond to prediction errors (on hidden states $x$ or hidden causes $\nu$). These are weighted by their respective precision vectors $\Pi^{(i)}$ in the update scheme. %
Equation~\ref{eq:3} can be derived fairly easily by computing the free energy
for the hierarchical model in Equation~\ref{eq:2} and inserting its gradients
into Equation~\ref{eq:1}. This gives a relatively simple update scheme, in
which conditional expectations are driven by a mixture of prediction
errors, where prediction errors are defined by the equations of the
generative model.

It is difficult to overstate the generality and importance of Equation~\ref{eq:3} -- its solutions grandfather nearly every known statistical
estimation scheme, under parametric assumptions about additive noise~\citep{Friston08b}. These range from ordinary least squares to advanced
variational deconvolution schemes. In this form, one can see clearly the
relationship between predictive coding and Kalman-Bucy filtering --
changes in conditional expectations comprise a prediction (first term)
plus a weighted mixture of prediction errors (remaining terms). The
weights play the role of a Kalman gain matrix and are based on the
gradients of the model functions and the precision of random
fluctuations.

In neural network terms, Equation~\ref{eq:3} says that error-units receive
predictions from the same hierarchical level and the level above.
Conversely, conditional expectations (encoded by the activity of state
units) are driven by prediction errors from the same level and the level
below. These constitute bottom-up and lateral messages that drive
conditional expectations towards a better prediction to reduce the
prediction error in the level below. This is the essence of recurrent
message passing between hierarchical levels to suppress free energy or
prediction error: see~\citep{Friston09b} for a more detailed
discussion. In neurobiological implementations of this scheme, the
sources of bottom-up prediction errors, in the cortex, are thought to be
superficial pyramidal cells that send forward connections to higher
cortical areas. Conversely, predictions are conveyed from deep pyramidal
cells by backward connections, to target (polysynaptically) the
superficial pyramidal cells encoding prediction error~\citep{Friston09b,Mumford92}. This defines an elementary circuit that may
be the basis of the layered organisation of the cortex~\citep{Bastos12}. Figure~\ref{fig:figure2} provides a schematic of the proposed message passing
among hierarchically deployed cortical areas.

\subsection{Action}
\label{sec:action}
In active inference, conditional expectations elicit behaviour by
sending predictions down the hierarchy to be unpacked into
proprioceptive predictions at the level of (pontine) cranial nerve
nuclei and spinal-cord. These engage classical reflex arcs to suppress
proprioceptive prediction errors and produce the predicted motor
trajectory

\begin{align}
\dot{a} =  - \partial_a F = -(\partial_a \tilde{\varepsilon}^{(1)}_\nu) \cdot \Pi^{(1)}_\nu \tilde{\varepsilon}^{(1)}_\nu  \label{eq:4} %
\end{align}%

The reduction of action to classical reflexes follows because the only
way that action can minimise free energy is to change sensory
(proprioceptive) prediction errors by changing sensory signals. This
highlights the tight relationship between action and perception; cf.,
the equilibrium point formulation of motor control~\citep{Feldman95}. In short, active inference can be regarded as equipping a
generalized predictive coding scheme with classical reflex arcs: see~\citep{Friston10a,Friston09c} for details. The actual
movements produced clearly depend upon (changing) top-down predictions
that can have a rich and complex structure. This scheme is consistent
with the physiology and anatomy of the oculomotor system (for a review
see~\citep{Ilg97,Krauzlis04}; although our goal here is not to identify
the role of each anatomical structure but rather to give a schematic
proof-of-concept.

\subsection{Summary}

In summary, we have derived equations for the dynamics of perception and
action using a free energy formulation of adaptive (Bayes-optimal)
exchanges with the world and a generative model that is both generic and
biologically plausible. A technical treatment of the material above will
be found in~\citep{Friston10d}, which provides the details of the
generalized filtering used to produce the simulations in the next
section. Before looking at these simulations, we consider how delays can
be incorporated into this scheme.

\section{Active inference with sensorimotor delays}
\label{sec:delay}
If action and sensations were not subject to delays, one could integrate
(solve) Equation~\ref{eq:1} directly; however, in the presence of sensory and
motor delays ($\tau_s$ and $\tau_a$ respectively) Equation~\ref{eq:1} becomes a (stochastic and non-linear) delay differential equation because
$\tilde{s}(t) = \bm{\tilde{s}}(t - \tau_s)$ and $a(t) = \bm{a}(t + \tau_a)$. In other words, the agent receives sensations from (sees) the past,
whilst emitting motor signals that will be enacted in the future (we
will only consider delays from the sensory and motor sub-systems and
neglect delays between neuronal systems in this paper).

To finesse the integration of these delay differential equations one can
exploit their formulation in generalized coordinates: By taking linear
mixtures of generalized motion one can easily map from the present to
the future, using the matrix operators:

\begin{align}%
 T(\tau) = \exp(\tau \mathcal{D})  &=  \left[\begin{array}{cccc}1 & \frac{1}{1!}\tau & \frac{1}{2!}\tau^2 & \ldots \\0 & 1 & \frac{1}{1!}\tau & \ldots \\0 & 0 & 1 & \ddots \\0 & 0 & 0 & \ddots\end{array}\right]  \label{eq:5} \\
 \text{ with }  \mathcal{D} &= \left[\begin{array}{cccc}0 & 1 & 0 & 0 \\0 & 0 & 1 & 0 \\0 & 0 & 0 & \ddots \\0 & 0 & 0 & 0\end{array}\right] \nonumber%
\end{align}%

The first differential operator simply returns the generalized motion $\mathcal{D}  \tilde{x}(t) = \tilde{x}^{\prime}(t)$ while the second delay operator produces generalized states in the
future $T(\tau)   \tilde{x}(t) = \tilde{x}(t+\tau)$ (we define delays as positive by convention). Note that shifting forward and backwards by the same amount of time produces the identity operator $T(\tau)  T(-\tau) = I$ and that, more generally, $T(\tau_1)  T(\tau_2) = T(\tau_1 + \tau_2)$.

These delay operators are simple to implement computationally (and
neurobiologically) and allow an agent to finesse the delayed coupling
above by replacing (delayed) sensory signals with future input
$\tilde{s}(t)=T(\tau_s)\bm{\tilde{s}}(t-\tau_s) = \bm{\tilde{s}}(t)$ for subsequent action and perception. Alternatively, one
can regard this compensation for sensory delays as attempting to predict
the past (see below). Generalized coordinates allow the representation
of the trajectory of a given variable at any time (that is, its
evolution in the near past and present) and thus allow its projection
into the future or past. Generalized representations are more extensive
than `snapshots' at a particular time and enable the agent to anticipate
the future (of delayed sensory trajectories) and represent hidden states
in real time -- that is, representations that are synchronised with the
external events. In terms of motor delays, the agent can replace its
internal motor signals with action in the future
$\bm{a}(t) = T(\tau_a) a(t-\tau_a) = a(t) $,
such that when action signals reach the periphery they correspond to the
action encoded centrally. These substitutions allow us to express action
and perception in Equation~\ref{eq:1}
as\footnote{We have a made a slight approximation here because $T(\tau_a) \tilde{\mu}(t-\bm{\tau}_a) = T(\tau_a-\bm{\tau}_a) \tilde{\mu}(t)$ when, and only when, the free energy gradients are zero and $\dot{\tilde{\mu}}(t) = \mathcal{D} \tilde{\mu}(t)$. Under the assumption that the perceptual destruction of these gradients is fast, in relation to action, this can
be regarded as an adiabatic approximation.}:

\begin{align}
\bm{\dot{a}}(t) &= -\partial_a F(T(\tau_a) T(\tau_s)  \bm{\tilde{s}}(t-\bm{\tau}_s-\bm{\tau}_a), T(\tau_a) \tilde{\mu}(t-\bm{\tau}_a))\nonumber  \\%
                &= -\partial_a F(T(\tau_s-\bm{\tau}_s+\tau_a-\bm{\tau}_a)  \bm{\tilde{s}}(t), T(\tau_a-\bm{\tau}_a) \tilde{\mu}(t))\nonumber  \\%
\label{eq:6} \\%
\dot{\tilde{\mu}}(t) &= \mathcal{D}\tilde{\mu}(t)-\partial_{\tilde{\mu}} F(T(\tau_s) \bm{\tilde{s}}(t-\bm{\tau}_s), \tilde{\mu}(t))\nonumber  \\%
                &= \mathcal{D}\tilde{\mu}(t)-\partial_{\tilde{\mu}} F(T(\tau_s-\bm{\tau}_s) \bm{\tilde{s}}(t), \tilde{\mu}(t))\nonumber  %
\end{align}%

This equation distinguishes between true delays ($\bm{\tau}$) and those assumed by the agent ($\tau$). 
When the two are the same, the delay operators  $T(\tau-\bm{\tau})=I: \tau=\bm{\tau}$ 
become identity matrices and Equation~\ref{eq:6} reduces to Equation~\ref{eq:1}. When
the two differ, Equation~\ref{eq:6} permits the simulation of a system with
uncompensated delays. Notice how the dynamics of action in the first
differential equation are driven by a gradient descent on the free
energy of sensations with composite sensory and motor delays. In other
words, action in the real world depends upon sensory states generated
$\bm{\tau}_s+\bm{\tau}_a$ in the past.

One can now solve Equation~\ref{eq:6} to simulate active inference, with or
without compensation for sensorimotor delays. We use a standard local
linearisation scheme for this integration~\citep{Ozaki92}, where delays
enter at the point at which sensory prediction error is computed and
when it drives action: from Equations~\ref{eq:3} and~\ref{eq:4}:

\begin{align}
\tilde{\varepsilon}^{(1)}_\nu &= T(\tau_s)  \bm{\tilde{s}}(t-\bm{\tau}_s) - \tilde{g}^{(1)}(\tilde{\mu}^{(1)}_x, \tilde{\mu}^{(1)}_\nu)  \nonumber \\%
                              &= T(\tau_s-\bm{\tau}_s)  \bm{\tilde{s}}(t) - \tilde{g}^{(1)}(\tilde{\mu}^{(1)}_x, \tilde{\mu}^{(1)}_\nu) \nonumber  \\%
\label{eq:7} \\%
\bm{\dot{a}}(t) &= -(\partial_a \tilde{\varepsilon}^{(1)}_\nu) \cdot \Pi^{(1)}_\nu T(\tau_a)  \tilde{\varepsilon}^{(1)}_\nu (t-\bm{\tau}_a) \nonumber  \\%
                &= -(\partial_a \tilde{\varepsilon}^{(1)}_\nu) \cdot \Pi^{(1)}_\nu T(\tau_a-\bm{\tau}_a)  \tilde{\varepsilon}^{(1)}_\nu (t) \nonumber  %
\end{align}%

Equation~\ref{eq:7} means that perfect (errorless) prediction requires
$T(\tau_s)  \bm{\tilde{s}}(t-\bm{\tau}_s) = \tilde{g}^{(1)}(\tilde{\mu}^{(1)}_x, \tilde{\mu}^{(1)}_\nu)$. In other words, errorless prediction means that the agent is effectively
predicting the future projection of the past. Note again the dependency
of action, via prediction errors, on sensory states
$\bm{\tau}_s+\bm{\tau}_a$ in the past. See Appendix 3 for further details of the integration scheme used in the simulations below.

\subsection{Summary}

This section has considered how the differential equations describing
changes in action and internal (representational) states can be finessed
to accommodate sensorimotor delays. This is relatively straightforward
-- in the context of generalized schemes -- using delay operators that
take mixtures of generalized motion to project states into the future or
past. Sensory delays can be (internally) simulated and corrected by
applying delays to sensory input producing sensory prediction error,
while motor delays can be simulated and corrected by applying delays to
sensory prediction error producing action. Neurobiologically, the
application of delay operators just means changing synaptic connection
strengths to take different mixtures of generalized sensations and their
prediction errors. We will now use these operators to look at the
effects of sensorimotor delays with and without compensation.

\section{Results: pursuit initiation}
\label{sec:pursuit}
This section focuses on the consequences of sensory delays, motor delays
and their combination -- in the context of pursuit initiation -- using
perhaps the simplest generative model for active inference. Our purpose
is to illustrate the difficulties in oculomotor control that are
incurred by delays and how these difficulties dissolve when delays are
accommodated during active inference. We start with a description of the
generative model and demonstrate its behaviour when delays are
compensated. We then use this normal behaviour as a reference to look at
failures of pursuit initiation induced by delays. In this section,
responses to a single sweep of rightward motion are used to illustrate
basic responses. In the next section, we consider pursuit of sinusoidal
motion (with abrupt onsets) and the implications for generative models
that may be used by the brain.

\subsection{Generative model of pursuit initiation}

The generative model for pursuit initiation used here is very simple and
is based upon the prior belief that the centre of gaze is attracted to
the target location. The processes generating sensory inputs and the
associated generative model can be expressed as follows:

\begin{align}
\bm{s} &=  \left[\begin{array}{c} \bm{s}_o \\ \bm{s}_t \end{array}\right] =  \left[\begin{array}{c} \bm{x}_o \\  \bm{x}_t - \bm{x}_o \end{array}\right] + \bm{\omega}^{(1)}_\nu \nonumber \\%
\bm{\dot{x}} &=  \left[\begin{array}{c} \bm{\dot{x}}_o \\ \bm{\dot{x}}_t \end{array}\right] =  \left[\begin{array}{c} \frac{1}{t_a} a-\frac{1}{t_o}\bm{x}_o \\  \frac{1}{t_m}(\bm{\nu}^{(1)} -  \bm{x}_t) \end{array}\right] + \bm{\omega}^{(1)}_x \nonumber \\%
 \label{eq:8} \\%
s &=  \left[\begin{array}{c} s_o \\ s_t \end{array}\right] =  \left[\begin{array}{c} x_o \\  x_t - x_o \end{array}\right] + \omega^{(1)}_\nu \nonumber \\%
\dot{x} &=  \left[\begin{array}{c} \dot{x}_o \\ \dot{x}_t \end{array}\right] =  \left[\begin{array}{c} \frac{1}{t_s}(x_t-x_o) \\  \frac{1}{t_m}(\nu^{(1)} -  x_t) \end{array}\right] + \omega^{(1)}_x \nonumber \\  %
\nu^{(1)} &= \omega^{(2)}_x \nonumber %
\end{align}%

The first pair of equations corresponds to a noisy sensory mapping from hidden states and the equations of motion for states in the real world. These pertain to real-world variables representing the position of the target and of the eye (in boldface). The remaining equations constitute the generative model of how sensations are generated using the form of Equation~\ref{eq:2}. These define the free energy in Equation~\ref{eq:1} -- and specify behaviour under active inference. The variables constitute the first layer of the hierarchical model (see Equation~\ref{eq:2}, but for simplicity, we have written $\bm{x}$ instead of $\bm{x}^{(1)}$ and ${x}$ instead of ${x}^{(1)}$.

The real-world provides sensory input in two modalities: proprioceptive input from cranial nerve nuclei reports the angular displacement of the eye $\bm{s}_o \in \mathbb{R}^2$ and corresponds to the centre of gaze. %
Note that, using the approximation of relatively small displacements, we use Cartesian coordinates to follow previous treatments e.g~\citep{Friston10d}. However visual space is better described by bounded polar coordinates, and treatments of large eye movements should account for this. %
Exteroceptive (retinal) input reports the angular position of a target in a retinal (intrinsic) frame of reference $\bm{s}_t \in \mathbb{R}^2$. The indices $o$ and $t$ thus refer to states of the oculomotor system or of the target, respectively. Note that $\bm{s}_t $ is just the difference between the centre of gaze and target location in an extrinsic frame of reference $\bm{x}_t - \bm{x}_o$. In this paper, we are modelling the online inference of target position, and we are ignoring the problem of how the causal structure of the environment is learned. We simply assume that this structure has already been learned accurately, and therefore the dynamics of the real-world and the generative model are the same. Clearly, this model of visual processing is an enormous simplification: we are assuming that place coded spatial information can be summarised in terms of displacement vectors. However, more realistic simulations -- using a set of retinotopic inputs with classical receptive fields covering visual space -- produce virtually the same results. We will use more realistic models in future publications that deal with smooth pursuit and visual occlusion. Here, we use the simpler formulation to focus on delays and the different sorts of generative models that can provide top-down or extra-retinal constraints on visual motion processing.

The hidden states of this model comprise the true, real-world oculomotor displacement ($\bm{x}_o \in \mathbb{R}^2$) and target location ($\bm{x}_t \in \mathbb{R}^2$). The units of angular displacement are arbitrary, but parameters are tuned to correspond to a small displacement of 4 degrees of visual angle for one arbitrary unit (that is approximately 4 times the width of a thumb's nail at arm's length). The oculomotor state is driven by action with a time constant of $t_a=64~\ms\ $ and decays (slowly) to zero through damping, with a time constant of $t_o = 512~\ms$. The target location is perturbed by hidden causes $\bm{x}_t \in \mathbb{R}^2$ that describe the location to which the target is drawn, with a time
constant of $t_m=16~\ms$. In this paper, the random fluctuations on sensory input and on the motion of hidden states are
very small, with a log precision of 16. In other words, the random fluctuations have a variance of $\exp(-16)$. This completes our description of the process generating sensory information; in which hidden causes force the motion of a target location and action forces oculomotor states. Target location and oculomotor states are combined to produce sensory information about the target in an intrinsic frame of reference.

The generative model has exactly the same form as the generative process
but with one important exception: there is no action and the motion of
the hidden oculomotor states is driven by the displacement between the
target location and the central gaze (with a time constant of $t_s=32~\ms$). In other words, the agent believes that its gaze will be
attracted to the location of the target, which, itself, is being driven
by some unknown exogenous force or hidden cause. The log-precisions on
the random fluctuations in the generative model were four, unless stated
otherwise. This means that uncertainty about sensory input, (motion of)
hidden states and causes were roughly equivalent.

Having specified the generative process and model, we can now solve the
active inference scheme in Equation~\ref{eq:1} and examine its behaviour.
Sensorimotor delays are implemented in the message passing from the
generative process to the generative model. This generative model
produces pursuit initiation because it embodies prior beliefs that the
centre of gaze will follow the target location. This pursuit initiation
rests on conditional expectations about the target location in extrinsic
coordinates and the state of the oculomotor plant, where the location is
driven by hidden causes that also have to be inferred.

The generative model described in this section provides the equations required to simulate active inference using the formalism of the previous section. In short, we now consider the generative model that defines the variational free energy and (Bayes) optimal active inference.

\subsection{Simulations}

\begin{figure}
 \centerline{%
 \includegraphics[width=.95\columnwidth, clip, trim = 7cm 3cm 7cm 2cm]{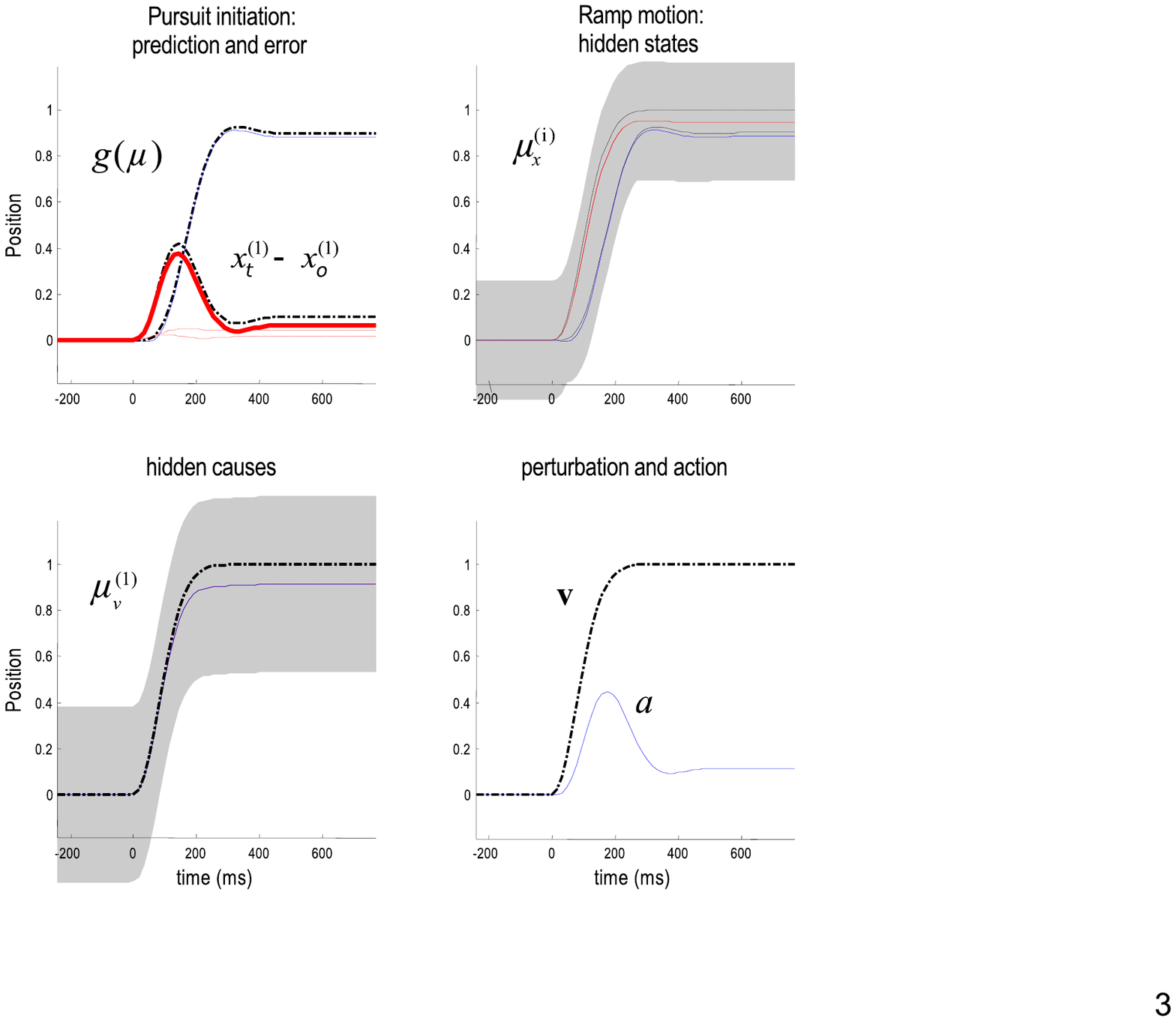}
 }%
\caption{This figure reports the conditional estimates of
hidden states and causes during the simulation of pursuit initiation,
using a single rightward (positive) sweep of a visual target, while
compensating for sensory motor delays. We will use the format of this
figure in subsequent figures: the upper left panel shows the predicted
sensory input (coloured lines) and sensory prediction errors (dotted red
lines) along with the true values (broken black lines). Here, we see
horizontal excursions of oculomotor angle (upper lines) and the angular
position of the target in an intrinsic frame of reference (lower lines).
This is effectively the distance of the target from the centre of gaze
and reports the spatial lag of the target that is being followed (solid
red line). One can see clearly the initial displacement of the target
that is suppressed after a few hundred milliseconds. The sensory
predictions are based upon the conditional expectations of hidden
oculomotor (blue line) and target (red line) angular displacements shown
on the upper right. The grey regions correspond to 90\% Bayesian
confidence intervals and the broken lines show the true values of these
hidden states. One can see the motion that elicits following responses
and the oculomotor excursion that follows with a short delay of about 64~\ms . The hidden cause of these displacements is shown with its
conditional expectation on the lower left. The true cause and action are
shown on the lower right. The action (blue line) is responsible for
oculomotor displacements and is driven by the proprioceptive prediction
errors.}%
\label{fig:figure3}
\end{figure}

\begin{figure}
 \centerline{%
 \includegraphics[width=.95\columnwidth, clip, trim = 7.cm .5cm 6.1cm .4cm, page=4]{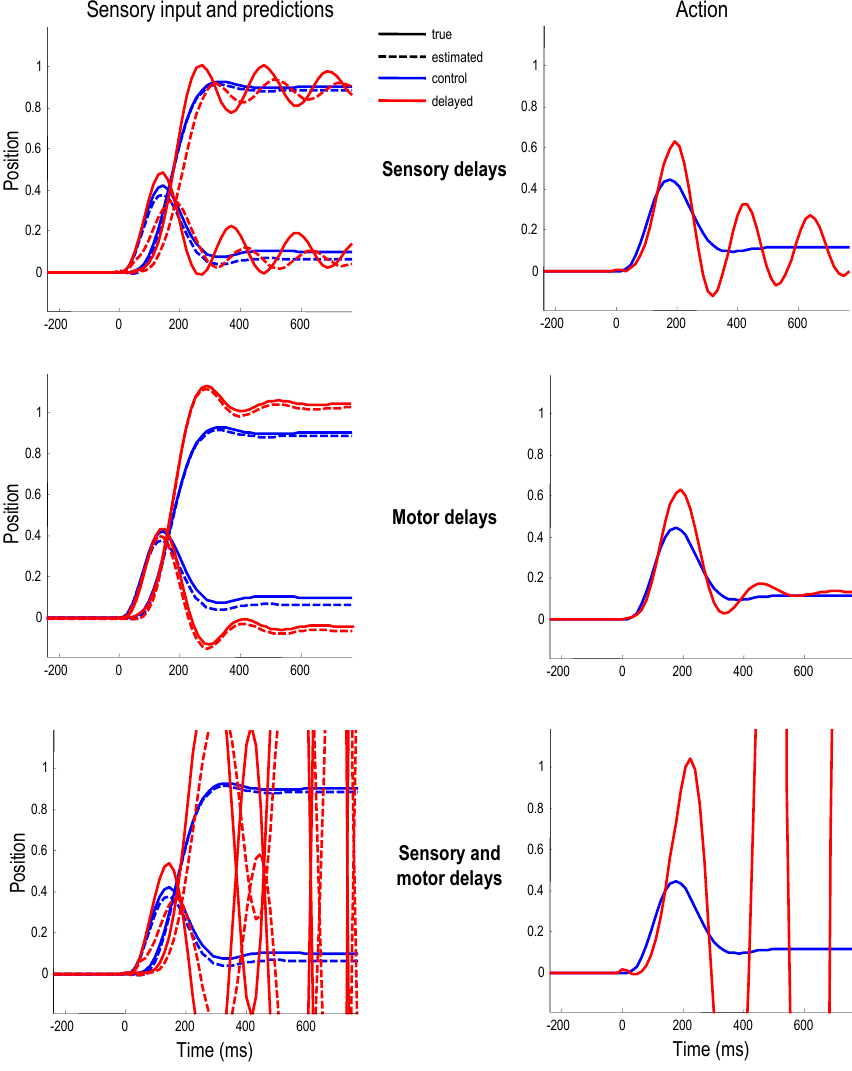} 
}%
\caption{This figure illustrates the effects of sensorimotor
delays on pursuit initiation (red lines) in relation to compensated
(optimal) active inference -- as shown in the previous figure (blue
lines). The left panels show the true (solid lines) and estimated
sensory input (dotted lines), while action is shown in the right panels.
Under pure sensory delays (top row), one can see clearly the delay in
sensory predictions, in relation to the true inputs. The thicker (solid
and dotted) red lines correspond respectively to (true and predicted)
proprioceptive input, reflecting oculomotor displacement. The middle row
shows the equivalent results with pure motor delays and the lower row
presents the results with combined sensorimotor delays. Of note here is
the failure of optimal control with oscillatory fluctuations in
oculomotor trajectories, which become unstable under combined
sensorimotor delays.}%
\label{fig:figure4}
\end{figure}
All simulations were performed with a time bin of 16ms and we report results in milliseconds. All results were replicated with different time bins (16ms, 8ms, 4ms, 2ms, 1ms) with minimal changes to the results. %
Figure~\ref{fig:figure3} reports the conditional estimates of hidden states and causes
during the simulation of pursuit initiation, using a simple rightward
sweep of a visual target and compensating for sensorimotor delays:
$\tau_s = \bm{\tau}_s $ and
$\tau_a = \bm{\tau}_a$.
This compensation is effectively the same as simulating responses in the
absence of delays -- because the delay operators reduce to the identity
matrix. Target motion was induced using a hidden cause that was a ramp
function of post-stimulus time. Note that ramp stimuli are often used in
psychophysics, and this generative model -- using velocity in place of
position --- produces the same results in velocity space. Indeed, most
models, such as~\citep{Robinson86} or~\citep{Krauzlis89},
focus on modelling velocity responses. We choose to model the tracking
of position for two reasons: First, it is easy to generalise position
results to velocity using generalized coordinates of motion. Second,
positional errors can induce slow eye movements~\citep{Kowler79,Wyatt81} and we hoped to accommodate this in the
model. If we assume that the units of angular displacement are 4 degrees of visual angle, then
the resulting peak motion corresponds to about 20 degrees per second.

The upper left panel shows the predicted sensory input (coloured lines)
and sensory prediction errors (dotted red lines) along with the true
values (broken black lines). Here, we see horizontal excursions of
oculomotor angle (upper lines) and the angular position of the target in
an intrinsic frame of reference (lower lines). This is effectively the
distance of the target from the centre of gaze and reports the spatial
lag of the target that is being followed (solid red line). One can see
clearly an initial retinal displacement of the target that is suppressed
after approximately $20~\ms $. This effect confirms that the visual representation of target position is predictive
and that the presentation of a smooth predictable versus an
unpredictable target would induce a lag between their relative
positional estimates, as is evidenced in the \emph{flash-lag
effect}~\citep{Nijhawan94}.

The sensory predictions are based upon the conditional expectations of
hidden oculomotor (blue line) and target (red line) angular
displacements shown on the upper right. The grey regions correspond to
90\% Bayesian confidence intervals and the broken lines show the true
values. One can see clearly the motion that elicits pursuit initiation
responses, where the oculomotor excursion follows with a short delay of
about $64~\ms$. The hidden cause of these
displacements is shown with its conditional expectation on the lower
left. The true cause and action are shown on the lower right. The action
(blue line) is responsible for oculomotor displacements and is driven by
proprioceptive prediction errors. %
Action does not return to zero because the sweep is maintained at an eccentric position during this simulation. This eye position slightly undershoots the target position: it is held at around 95\% of the target eccentricity in the upper right panel. Note that this corresponds roughly to the steady-state gain observed in behavioural data, which was modelled explicitly by~\citep{Robinson86}. %
For our purposes, these simulations can be regarded as
Bayes optimal solutions to the pursuit initiation problem, in which
sensorimotor delays have been accommodated (discounted) via absorption
into the generative model. We can now examine the performance in the
absence of compensation and see how sensory and motor delays interact to
confound pursuit initiation:

The above simulations were repeated with uncompensated sensory delays ($\tau_s =0~\ms$ and $\bm{\tau}_s=32~\ms$), uncompensated motor delays ($\tau_a =0~\ms$ and $\bm{\tau}_a=32~\ms$) and combined sensorimotor delays of $64~\ms$
($\tau_a =\tau_s =0~\ms$ and $\bm{\tau}_a=\bm{\tau}_s=32~\ms$).
To quantify behaviour, we focus on the sensory input and underlying
action. The position of the target in intrinsic coordinates corresponds
to spatial lag and usefully quantifies pursuit initiation performance.
Figure~\ref{fig:figure4} shows the results of these three simulations (red lines) in
relation to the compensated (optimal) active inference shown in the
previous figure (blue lines). True sensory input corresponds to solid
lines and its conditional predictions to dotted lines. The left panels
show the true and predicted sensory input, while action is shown in the
right panels. Under pure sensory delays (top row) one can see the delay
in sensory predictions, in relation to the true inputs. The thicker
(solid and dotted) red lines correspond respectively to (true and
predicted) proprioceptive input, reflecting oculomotor displacement.
Crucially, in contrast to optimal control, there are oscillatory fluctuations in oculomotor displacement and the retinotopic location of the target that persists even after the target is stationary. These fluctuations are similar to the oscillations elicited by adding an artificial feed-back delay ~\citep{Goldreich92}. Here, the fluctuations are caused by damped oscillations in action due to, and only to, sensory proprioceptive and exteroceptive delays. These become unstable (increasing in their amplitude) when the predicted value oscillates in counter-phase with the real value. %
Similar
oscillations are observed with pure motor delays (middle row). However,
here there is no temporal lag between the true and predicted sensations
(solid versus dashed lines). Furthermore, there is no apparent delay in
action -- action appears to be emitted for longer, reaching higher
amplitudes. In fact, action is delayed but the delay is obscured by the
increase in the amplitude of action -- that is induced by greater
proprioceptive prediction errors. If we now combine both sensory and
motor delays, we see a catastrophic failure of oculomotor tracking
(lower row). With combined sensorimotor delays the pursuit initiation
becomes unstable, with exponentially increasing oscillations as action
over-compensates for delay-dependent errors.

In effect, the active inference scheme has undergone a phase transition
from a stable to an unstable fixed point. We have illustrated this
bifurcation by increasing sensorimotor delays under a fixed motor
precision or gain in Equation~\ref{eq:7}. The results in Figure~\ref{fig:figure4} used a motor
gain with a log precision of 2.5. We chose this value because it
produced stable responses with sensory or motor delays alone and
unstable dynamics with combined delays. These results illustrate the
profound and deleterious effects of sensorimotor delays on simple
pursuit initiation, using biologically plausible values -- namely
sensorimotor delays of 64~\ms\ and a target velocity of about 16 degrees per
second. This also illustrates the necessity of compensation for these
delays so that the system can achieve a more robust and stable response.
One would anticipate, in the face of such failures, real subjects would
engage interceptive saccades to catch the target, of the sort seen in
schizophrenic patients~\citep{Levy93}. In the remainder of this
paper, we will concentrate on the nature of pursuit initiation and
smooth pursuit with compensated sensorimotor delays, using a reasonably
high motor gain with a log precision of four.

\subsection{Pursuit initiation and visual contrast}

\begin{figure}
 \centerline{%
 \includegraphics[width=.8\columnwidth, clip, trim = 7cm 1cm 7cm .4cm, page=5]{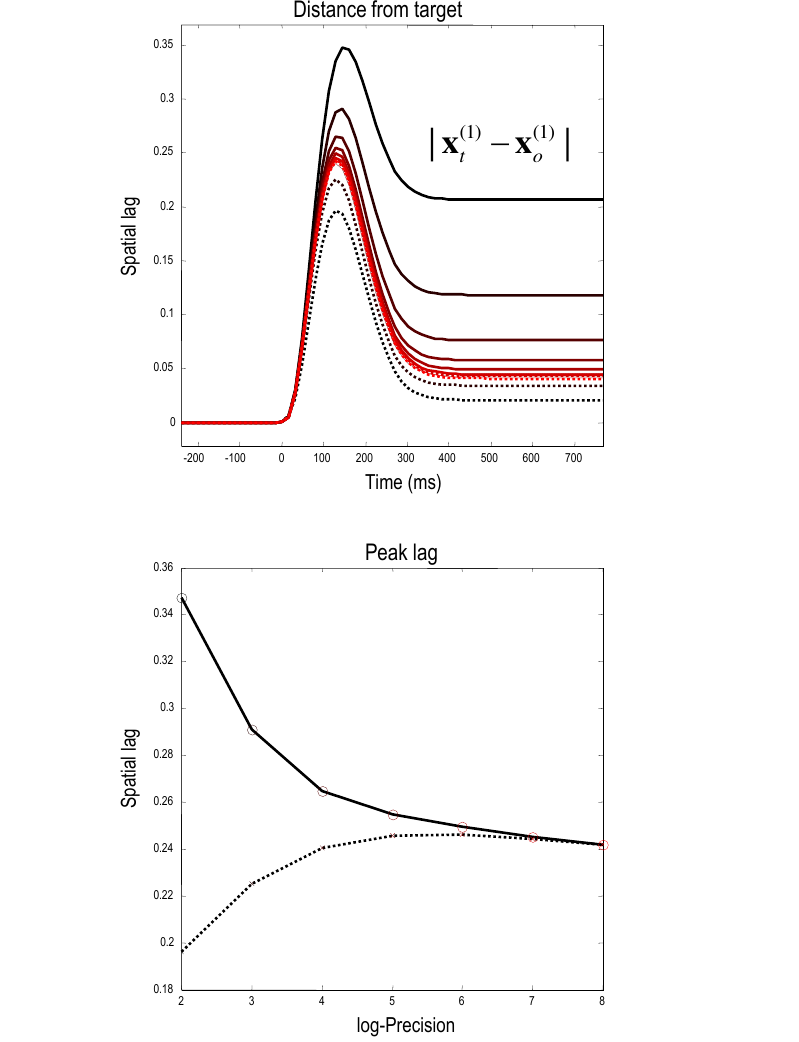} 
}%
\caption{ This figure reports the spatial lag (the displacement
of the target from the centre of gaze) as a function of contrast (log
precision of exteroceptive sensory input). The upper panel shows the
true (solid lines) and predicted (dotted lines) spatial lag as a
function of peristimulus time for different log precisions, ranging from
two (black lines) to eight (red lines). The peak lags are plotted in the
lower panel as a function of visual contrast or log precision. These
results show how the perceived lag increases with contrast, while the
true lag decreases in accord with empirical observations.}%
\label{fig:figure5}
\end{figure}

Before turning to more realistic generative models of smooth pursuit, we
consider the empirical phenomena in which following responses to the
onset of target movement are suppressed by reducing the visual contrast
of the target~\citep{Thompson82}. In simulations of this sort, visual
contrast is modelled in terms of the precision of sensory information in
accord with Weber's law -- see~\citep{Feldman10a} for details.
Contrast-dependent effects are easy to demonstrate in the context of
active inference. Figure~\ref{fig:figure5} shows the spatial lag -- the displacement in
intrinsic coordinates of the target from the centre of gaze depicted by
the solid red line in Figure~\ref{fig:figure3} -- as a function of contrast or
log-precision of exteroceptive sensory input. The upper panel shows the
true (solid lines) and predicted (dotted lines) spatial lag as a
function of peristimulus time for different log precisions, ranging from
two (low) to eight (high). The peak lags are plotted in the lower panel
as a function of visual contrast or log precision. Since estimation
error decreases as visual contrast increases, both curves converge,
leading to a decrease to zero of the prediction error. These results
show, in accord with empirical observations, how the spatial lag
(position error) increases with contrast~\citep{Arnold09}, while the
true lag decreases~\citep{Spering05}. A similar difference between perception and action was recently reported~\citep{Simoncini12}. The explanation for this
contrast--dependent behaviour is straightforward -- because pursuit
initiation is based upon proprioceptive prediction errors, it depends
upon precise sensory information. Reducing the precision of visual input
-- through reducing contrast -- increases uncertainty about visual
information (sensory estimation error) and places more weight on prior
beliefs and proprioceptive sensations. This reduces the perceived motion
of the target and reduces the amplitude of prediction errors driving
action.

\begin{figure}
 \centerline{%
 \includegraphics[width=.8\columnwidth, clip, trim = 5.5cm .5cm 6cm .4cm, page=6]{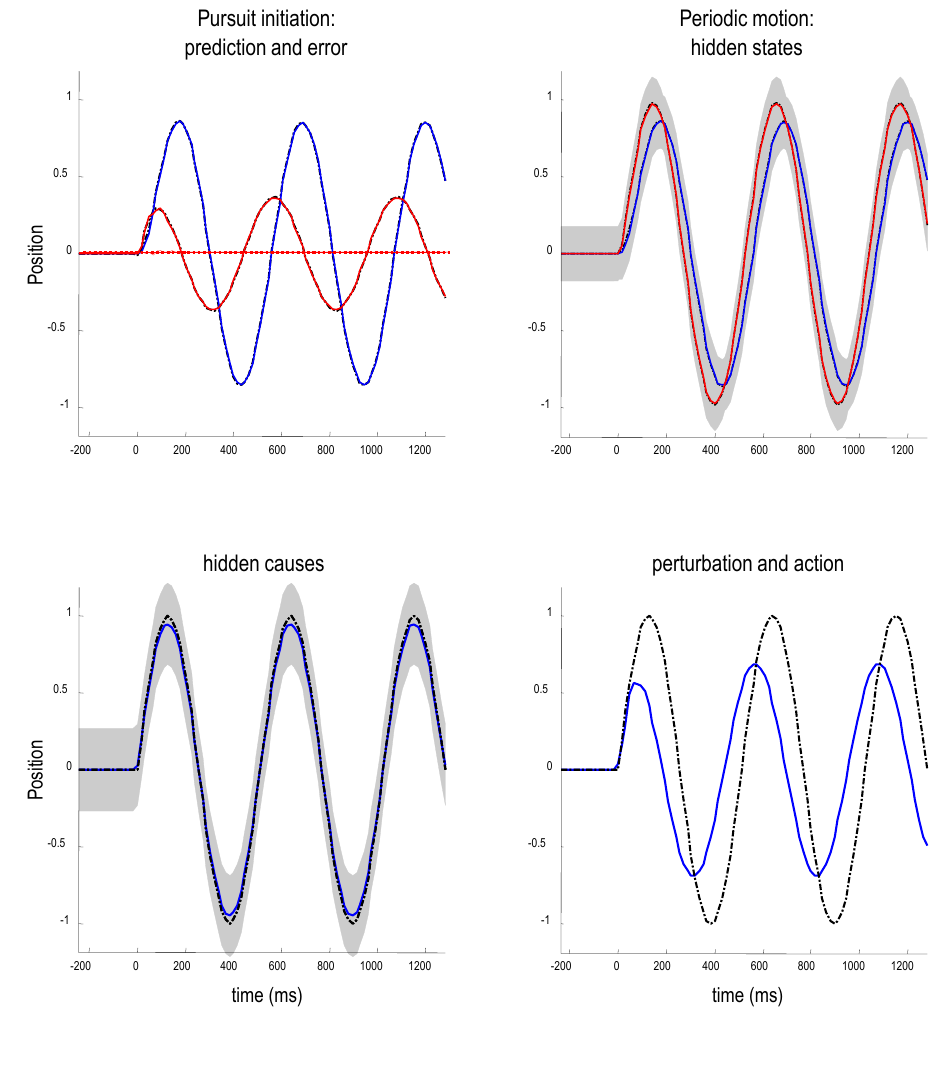} 
}%
\caption{This figure uses the same format as Figure~\ref{fig:figure3} -- the
only difference here is that the target motion is sinusoidal. The key
thing to take from this simulation is that the peak spatial lag at the
onset of the second cycle of target motion is greater than the peak lag
at the onset of the first. This is contrary to empirical predictions.}%
\label{fig:figure6}
\end{figure}

\subsection{Summary}

In this section, we have seen that sensorimotor delays can have profound
and deleterious effects on optimal oculomotor control. Here, optimal
control means Bayes optimal active inference, in which pursuit
initiation emerges spontaneously from prior beliefs about how a target
attracts the centre of gaze. These simulations demonstrate that it is
relatively easy to compensate for sensorimotor delays by exploiting
representations in generalized coordinates of motion. Furthermore, the
resulting scheme has some construct validity in relation to experimental
manipulations of the precision or contrast of visual information.
However, there are certain aspects of oculomotor tracking that suggest
the pursuit initiation model above is incomplete: when presented with
periodic target motion, the latency of motor gain (defined operationally
in terms of the target and oculomotor velocities) characteristically
reduces after the first cycle of target motion~\citep{Barnes00}.
This phenomenon cannot be reproduced by the pursuit initiation model
above:

Figure~\ref{fig:figure6} shows the responses of the pursuit initiation model to
sinusoidal motion using the same format as Figure~\ref{fig:figure3}. Here, the hidden
cause driving the target was a sine wave with a period of $512~\ms$ that started after $256~\ms$. If we focus on the
spatial lag (solid red line in the upper left panel), one can see that
the lag is actually greater after one period of motion than at the onset
of motion. This contrasts with empirical observations, which suggest
that the spatial lag should be smaller after the first cycle~\citep{Barnes00}. In the next section, we consider a more realistic generative
model that resolves this discrepancy and takes us from simple pursuit
initiation to smooth pursuit.

\section{Results: smooth pursuit}
\label{sec:spem}
In this section, we consider a slightly more realistic generative model
that replaces the prior beliefs about the target attracting the centre
of gaze with the belief that both the target and centre of gaze are
attracted by the same (fictive) location in visual space. This allows
pursuit initiation to anticipate the trajectory of the target and pursue
the target more accurately -- providing the trajectories are
sufficiently smooth. The idea behind this generative model is to account
for the improvements in tracking performance that are not possible at
the onset of motion and that are due to inference on smooth target
trajectories.

\subsection{Smooth pursuit model}

The smooth pursuit model considered in this paper rests on a
second-order generalisation of the pursuit initiation model of previous
section. Previously, we have considered the motion of the oculomotor
plant to be driven directly by action. This form of action can be
considered as an (adiabatic) solution to a proper second-order
formulation, in which action exerts a force and thereby changes the
angular acceleration of oculomotor displacement. This second-order
formulation can be expressed in terms of the following generative
process and model

\begin{align}
\bm{s} &=  \left[\begin{array}{c} \bm{s}_o \\ \bm{s}_t \end{array}\right] =  \left[\begin{array}{c} \bm{x}_o \\  \bm{x}_t - \bm{x}_o \end{array}\right] + \bm{\omega}^{(1)}_\nu \nonumber \\%
\bm{\dot{x}} &=  \left[\begin{array}{c} \bm{\dot{x}}_o \\ \bm{\dot{x}}^{\prime}_o \\ \bm{\dot{x}}_t \end{array}\right] =  \left[\begin{array}{c} \bm{x}^{\prime}_t \\ \frac{1}{t_a} a-\frac{1}{t_o}\bm{x}^{\prime}_o \\  \frac{1}{t_m}(\bm{\nu}^{(1)} -  \bm{x}_t) \end{array}\right] + \bm{\omega}^{(1)}_x \nonumber \\%
 \label{eq:9} \\%
s &=  \left[\begin{array}{c} s_o \\ s_t \end{array}\right] =  \left[\begin{array}{c} x_o \\  x_t - x_o \end{array}\right] + \omega^{(1)}_\nu \nonumber \\%
\dot{x} &=  \left[\begin{array}{c} \dot{x}_o \\ \dot{x}^{\prime}_o \\ \dot{x}_t \end{array}\right] =  \left[\begin{array}{c} {x}^{\prime}_t \\ \frac{1}{t_v}(\nu^{(1)}-x_o) - \frac{t_s}{t_v}{x}^{\prime}_o \\  \frac{1}{t_m}(\nu^{(1)} -  x_t)\end{array}\right] + \omega^{(1)}_x \nonumber \\  %
\nu^{(1)} &= \omega^{(2)}_\nu \nonumber %
\end{align}%

Here, the only thing that has changed is that we have introduced new
hidden states corresponding to oculomotor velocity
$\bm{x}^{\prime}_o \in \mathbb{R}^2$.
Action now changes the motion of the velocity (i.e., acceleration), as
opposed to the velocity directly. This difference is reflected in the
generative model but with one crucial addition -- the hidden oculomotor
state is not driven by the displacement between the \emph{target} and
the centre of gaze but by the displacement between the \emph{hidden
cause} and the centre of gaze. In other words, the hidden oculomotor
states are attracted by the hidden cause of target motion -- not the
target motion \emph{per se}. The idea here is that inference about the
trajectory of the hidden cause should enable an anticipatory
optimisation of pursuit initiation, provided these trajectories are
smooth -- hence a smooth pursuit model. Note that the equation of motion
in the oculomotor model
$\dot{x}_o = \frac{1}{t_s}(x_t-x_o)$
(see Equation~\ref{eq:8}) is the (adiabatic) solution to the equation used to
model smooth pursuit:
$\frac{1}{t_v}(\nu^{(1)}-x_o) - \frac{t_s}{t_v}{x}^{\prime}_o =0 $ when
$\nu^{(1)} =  x_t$ (see
Equation~\ref{eq:9}). As a result (and as confirmed by simulations) this model
behaved similarly for the sweep stimulus used in Figures~\ref{fig:figure3}, \ref{fig:figure4}, \ref{fig:figure5}. 

\subsection{Simulations}

\begin{figure}
 \centerline{%
 \includegraphics[width=.8\columnwidth, clip, trim = 6cm .5cm 6cm .4cm, page=7]{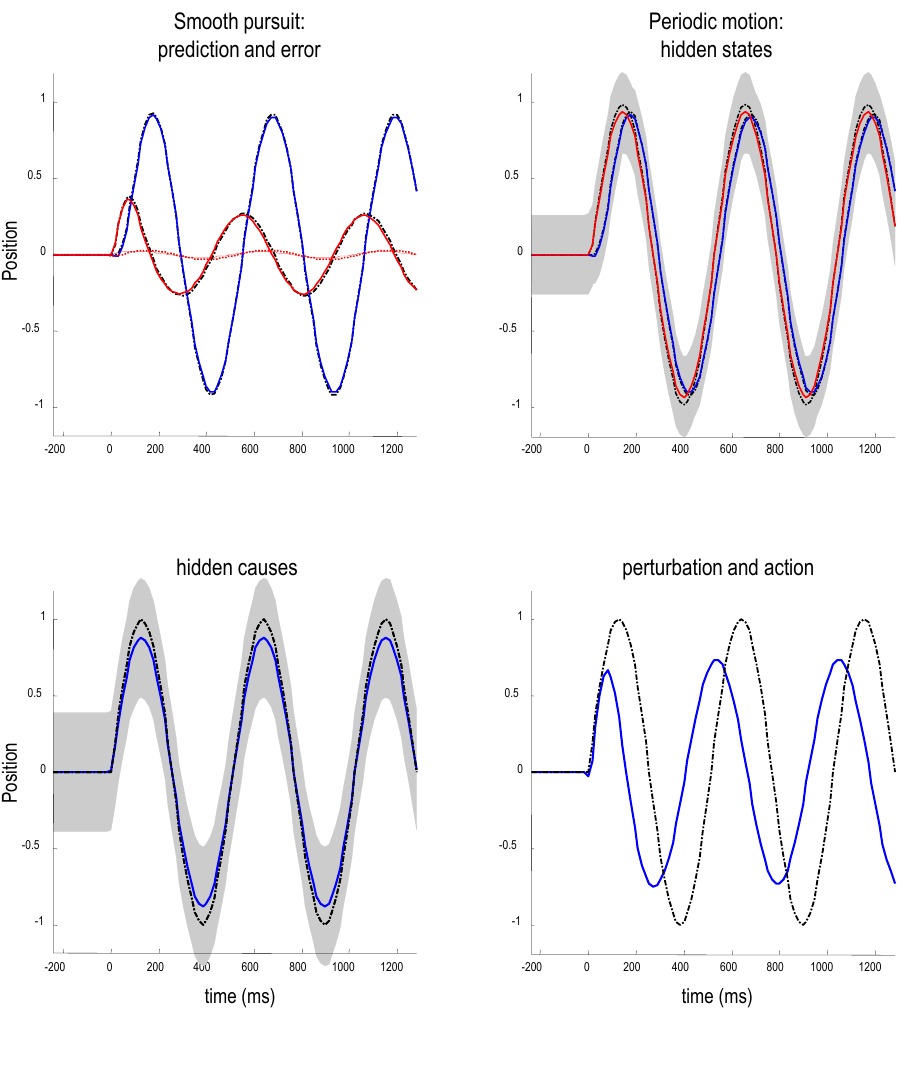} 
}%
\caption{This figure uses the same format as the previous
figure -- the only difference here is that we have replaced the pursuit
initiation model with a smooth pursuit model. In the smooth pursuit
model the centre of gaze is attracted by a hidden cause of target
motion, as opposed to the target \emph{per se}. Note that, in comparison
to the previous figure, the peak lag at the onset of the second cycle of
target motion is now smaller than at the onset to the first.}%
\label{fig:figure7}
\end{figure}

\begin{figure}
 \centerline{%
 \includegraphics[width=.8\columnwidth, clip, trim = 5.5cm .5cm 6.5cm .4cm, page=8]{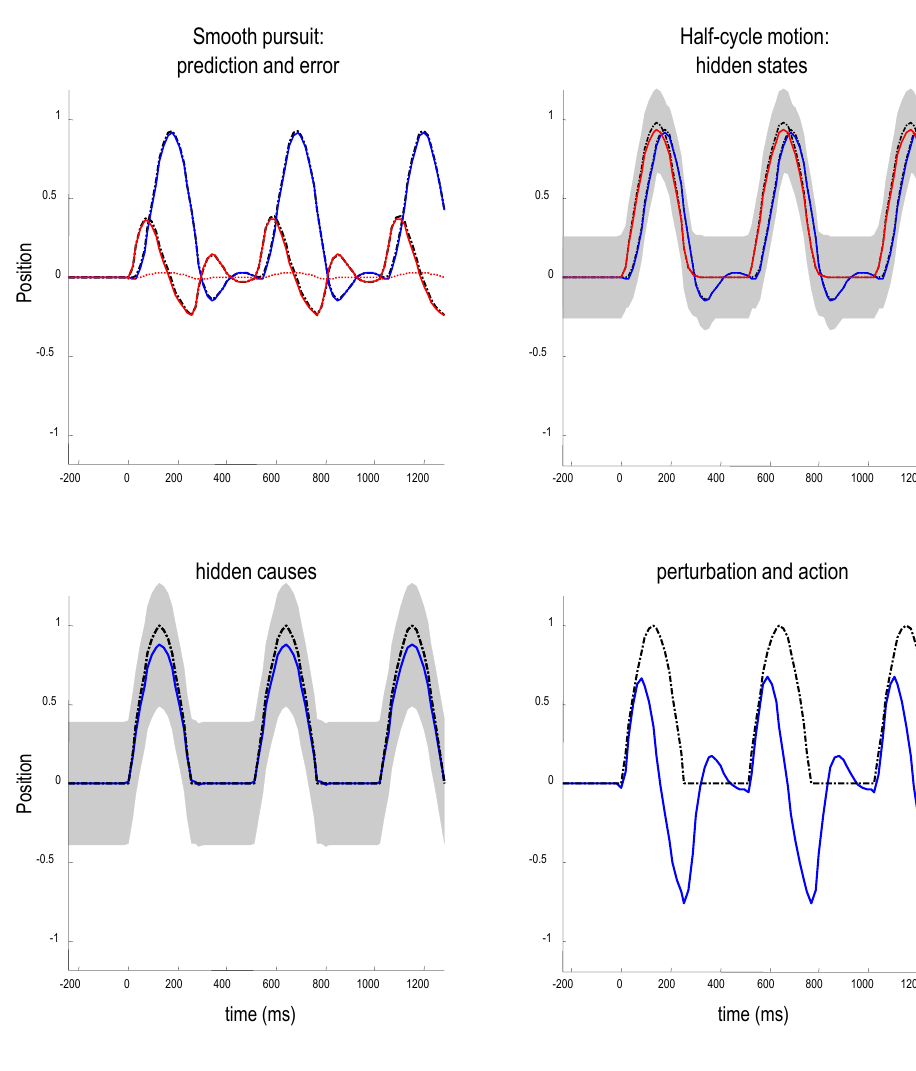} 
}%
\caption{This figure uses the same format as the previous
figure -- the only difference is that the target motion has been
rectified so that it is (approximately) hemi-sinusoidal. The thing to
note here is that the improved accuracy of the pursuit previously
apparent at the onset of the second cycle of motion has now disappeared
-- because active inference does not have access to the immediately
preceding trajectory. This failure of an anticipatory improvement in
tracking is contrary to empirical predictions.}%
\label{fig:figure8}
\end{figure}

We repeated the simulation reported in Figure~\ref{fig:figure6} using the smooth pursuit
generative model. The results of this simulation are shown in Figure~\ref{fig:figure7}
using the same format as Figure~\ref{fig:figure6}. The key difference -- in terms of
performance -- is that the peak spatial lag after one cycle of motion is
now less than the peak lag at the onset of motion. The response to the
sinusoid trajectory contrasts with simple pursuit initiation and is more
consistent with empirical observations. The true and expected hidden
states show that the oculomotor trajectory now follows the target
trajectory more accurately, particularly at the peaks of right and
leftward displacement. Interestingly, the amplitude of action has not
changed very much (compare Figures~\ref{fig:figure6} and~\ref{fig:figure7}, upper right panels).
However, action is initiated with a slightly shorter latency, which is
sufficient to account for the improved pursuit when informed by the
prior beliefs about the smooth trajectory of the target.

\subsection{Summary}

In summary, by simply replacing the target with the hidden cause of
target motion -- as the attractor of oculomotor trajectories -- we can
account for empirical observations of improved pursuit during periodic
target motion. In the context of active inference, this smooth
trajectory can only be recognised -- and used to inform action -- after
the onset of periodic motion. However, this smooth pursuit model still
fails to account for anticipatory effects that are not directly
available in sensory trajectories. Empirical observations suggest that
any systematic or regular structure in target motion can facilitate the
accuracy of smooth pursuit, even if this information is not represented
explicitly in target motion. A nice example of this rests on the use of
rectified periodic motion, in which only rightward target excursions are
presented. Experimentally, subjects can anticipate the periodic but
abrupt onset of motion, provided they recognise the underlying periodic
behaviour of the target. We can emulate this hemi-periodic motion by
thresholding the hidden cause to suppress leftward deflections. Figure~\ref{fig:figure8}
shows the results of simulating smooth pursuit using the same format as
Figure~\ref{fig:figure7}. The only difference here is that we replaced the sinusoidal
hidden cause
$\bm{\nu}(t) = \sin(2\pi f \cdot t)$ 
with
$\bm{\nu}(t)= \exp(4(\sin(2\pi f \cdot t)-1))$.
This essentially suppresses motion before rightward motion. This
suppression completely removes the benefit of smooth pursuit after a
cycle of motion -- compare Figures~\ref{fig:figure7} and~\ref{fig:figure8}. Here, the peak spatial lag
at the onset of the second cycle of motion is exactly the same as the
lag at the onset of motion; in other words, there is no apparent benefit
of modelling the hidden causes of motion in terms of pursuit accuracy.
This failure to model the anticipatory eye movements seen experimentally
leads us to consider a full hierarchical model for anticipatory pursuit.
\section{Results: anticipatory pursuit}
\label{sec:anticip}

This section presents a full hierarchical model of anticipatory smooth
pursuit eye movements that tries to account for anticipatory oculomotor
responses that are driven by extra-retinal beliefs about the periodic
behaviour of targets. This entails adding a hierarchical level to the
model that enables the agent to recognise and remember the latent
structure in target trajectories and suitably optimise its pursuit
movements -- which are illustrated here in terms of an improvement in
the accuracy of target following after the onset of rectified target
motion.

\subsection{Anticipatory pursuit}

The generative process used in these simulations is exactly the same as
in the above (smooth pursuit) scheme (see Equation~\ref{eq:9}); however, the
generative model of this process is equipped with an extra level in
place of the model for the hidden cause of target motion in the
generative model:

\begin{align}
s &=  \left[\begin{array}{c} s_o \\ s_t \end{array}\right] =  \left[\begin{array}{c} x_o \\  x_t - x_o \end{array}\right] + \omega^{(1)}_\nu \nonumber \\%
\dot{x} &=  \left[\begin{array}{c} \dot{x}_o \\ \dot{x}^{\prime}_o \\ \dot{x}_t \end{array}\right] =  \left[\begin{array}{c} {x}^{\prime}_t \\ \frac{1}{t_v}(\nu^{(1)}-x_o) - \frac{t_s}{t_v}{x}^{\prime}_o \\  \frac{1}{t_m}(\nu^{(1)} -  x_t) \end{array}\right] + \omega^{(1)}_x \nonumber \\  %
 \label{eq:10} \\%
\nu^{(1)} &=  \left[\begin{array}{c} \sigma(x^{(2)}_1 ) \\ 0 \end{array}\right] + \omega^{(2)}_\nu \nonumber \\%
\dot{x}^{(2)} &=  \left[\begin{array}{c} \dot{x}^{(2)}_1 \\ \dot{x}^{(2)}_2 \end{array}\right] =  \nu^{(2)} \left[\begin{array}{c} x^{(2)}_2 \\ -x^{(2)}_1 \end{array}\right] + \omega^{(2)}_x \nonumber \\  %
\nu^{(2)} &= \eta + \omega^{(3)}_\nu \nonumber %
\end{align}%

The first level of the generative model is exactly the same as above.
However, the hidden causes are now informed by the dynamics of hidden
states at the second level. These hidden states model underlying
periodic dynamics using a simple periodic attractor that produces
sinusoidal fluctuations of any amplitude or phase and a frequency that
is determined by a second level hidden cause with a prior expectation of a frequency of
$\eta$ (in \Hz).
It is somewhat similar to a control system model that attempted to
achieve zero-latency target tracking by fitting the trajectory to a
(known) periodic signal~\citep{Bahill83}. Our formulation
ensures a Bayes optimal estimate of periodic motion in terms posterior
beliefs about its frequency. In these simulations, we used a fixed Gaussian prior centred on the correct frequency with a period of $512~\ms$. This prior reproduces a typical experimental
setting in which the oscillatory nature of the trajectory is known, but
its amplitude and phase (onset) are unknown. Indeed, it has been shown
that anticipatory responses are cofounded when randomising the
inter-cycle interval~\citep{Becker85}. In principle, we could
have considered many other forms of generative model, such as models
with prior beliefs about continuous acceleration~\citep{Bennett10}.

As above, all the random fluctuations were assumed to have a log
precision of four. Crucially, the mapping between the second level
(latent) hidden states and the motion of first level hidden states
encoding trajectories in visual (extrinsic) space is nonlinear. This
means that latent periodic motion can be distorted in any arbitrary way.
Here, we use a soft thresholding function
$\sigma(x) = \exp(4(x-1))$
to suppress negative (rightward) excursions of the target to model
hemi-sinusoidal motion. This is the same function we used to generate
the motion in Figure~\ref{fig:figure8}. 
Note that if the precision of the noise at the second level falls to zero and there is no (precise) information at this level, the generative model assumes that the random fluctuations have an infinite variance. As a consequence, the prediction at the level below in the hierarchical model simplifies to $\nu^{(1)} = \omega^{(2)}_\nu $, and we recover Equation~\ref{eq:9} describing the smooth pursuit model. %
As a consequence this parameter tunes the relative strength of anticipatory modulation.

\begin{figure}
 \centerline{%
 \includegraphics[width=.8\columnwidth, clip, trim = 7cm 0cm 7cm 0cm, page=9]{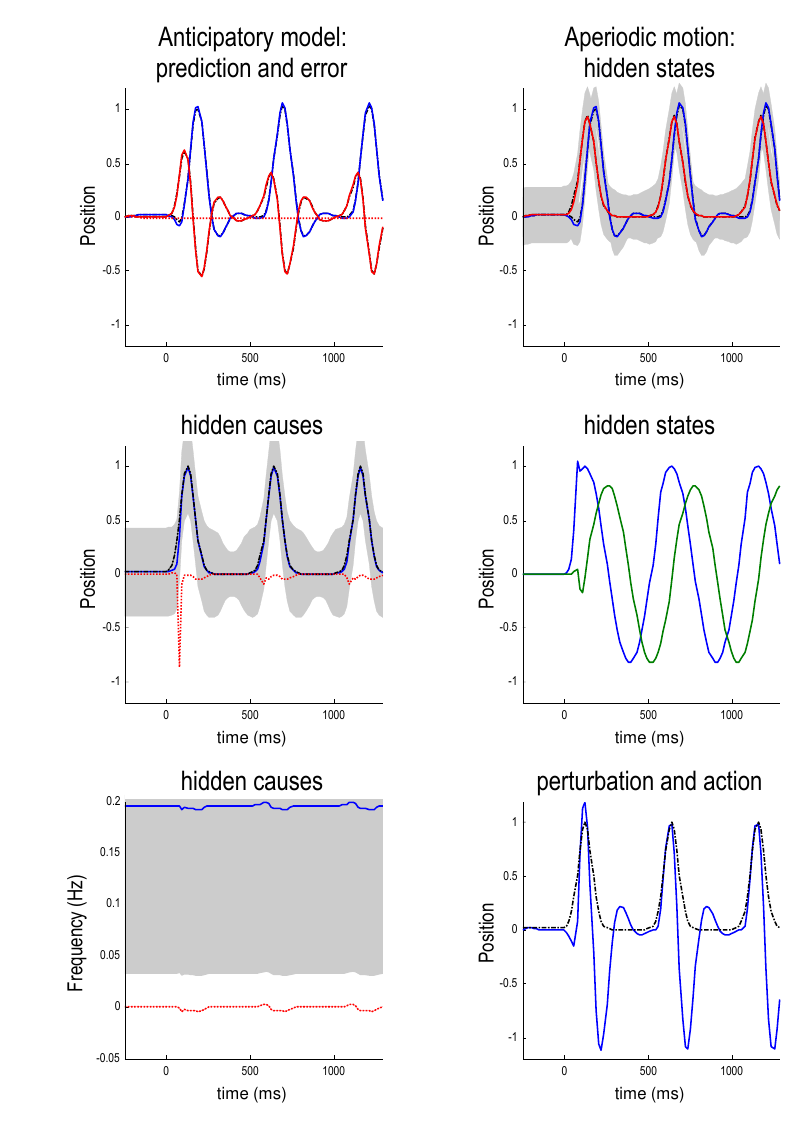} 
}%
\caption{This figure uses the same format as the previous
figure -- the only difference is that the generative model has been
equipped with a second hierarchical level that contains hidden states,
modelling latent periodic behaviour of the (hidden) causes of target
motion. With this addition, the improvement in pursuit accuracy apparent
at the onset of the second cycle of motion is reinstated. This is
because the model has an internal representation of latent causes of
target motion that can be called upon even when these causes are not
expressed explicitly in the target trajectory.}%
\label{fig:figure9}
\end{figure}

Figure~\ref{fig:figure9} shows the results of simulating active inference under this
anticipatory model, using the same format as Figure~\ref{fig:figure8}. However, there is
now an extra level of hidden states encoding latent periodic motion. It
can be seen that expectations about hidden states attain nonzero
amplitudes shortly after motion onset and are periodic thereafter. These
provide predictions about the onset of rightward motion after the first
(latent) cycle, enabling a more accurate oculomotor response. This is
evidenced by the reduction in the spatial lag at the onset of the second
cycle of motion, relative to the first (solid red lines on the upper
left). This improvement in accuracy should be compared to the previous
figure and reflects Bayes optimal anticipatory responses of the sort
observed empirically~\citep{Barnes00}. Further evidence of
anticipatory inference can be seen by examining the conditional
expectations about hidden causes at the second level. Note the
substantial reduction in prediction error on the hidden cause (dotted
red lines), when comparing the onset of the second cycle to the onset of
the first. This reflects the fact that the conditional expectations
about the hidden cause show a much-reduced latency at the onset of the
second cycle due to top-down conditional predictions provided by the
second level hidden states. This recurrent and hierarchically informed
inference provides the basis for anticipatory oculomotor control and may
be a useful metaphor for the hierarchical anatomy of the
visual-oculomotor system.

\subsection{Summary}

In conclusion, to account for anticipatory pursuit movements that are
not immediately available in target motion, one needs to equip
generative models with a hierarchal structure that can accommodate
latent dynamics -- that may or may not be expressed at the sensory
level. %
It is important to note that this model is a gross simplification of the complicated hierarchies that may exist in the brain. For instance, while some anticipation may be induced in smooth pursuit eye movements, some aspects, such as the aperture problem, may not be anticipated~\citep{Montagnini06}. In this model, the second level hidden causes are simply driven by prediction errors and assume a constant frequency. As a consequence, prior beliefs about frequency are modelled as stationary. In the real brain one might imagine that models of increasing hierarchical depth might allow for nonstationary frequencies and other dynamics -- that would better fit behavioural data. We have chosen to illustrate the basic ideas using a minimalistic example of anticipation in eye movements. %
Hierarchical extensions of this sort emphasise the distinction between visual motion processing and attending oculomotor control based purely upon retinal and proprioceptive input -- they emphasise extra-retinal processing that is informed by prior experience and beliefs about the latent causes of visual input. We will exploit this anticipatory smooth pursuit model in future work, where visual occluders
are used to disclose beliefs about latent motion.

\section{Discussion}
In this paper, we have considered optimal motor control in the context of pursuit initiation and anticipatory smooth pursuit. In particular, we have taken a Bayesian perspective on optimality and have simulated various aspects of eye movement control using predictive coding and active inference. This provides a solution to the problem of sensorimotor delays that reproduces the results of earlier solutions --- but using a neuronally plausible (predictive coding) scheme that has been applied to a whole range of perceptual, psychophysical, decision theoretic and motor control problems beyond oculomotor control.
Active inference depends upon a generative model of stimulus trajectories and
their active sampling through movement. This requires a careful
consideration of the generative models that might be embodied by the
visual-oculomotor system -- and the sorts of behaviours one would expect
to see under these models. The treatment in this paper distinguishes
between three levels of predictive coding with respect to oculomotor
control: the first is at the lowest level of sensorimotor message
passing between the sensorium and internal states representing the
causes of sensory signals. Here, we examined the potentially
catastrophic effects of sensorimotor delays and how they can easily
render oculomotor tracking inherently unstable. This problem can be
finessed -- in a relatively straightforward way -- by exploiting
representations in generalized coordinates of motion. These can be used
to offset both sensory and motor delays, using simple and
neurobiologically plausible mixtures of generalized motion. We then
motivated a model of smooth pursuit eye movements by noting that a
simple model of target following cannot account for the improvement in
visual tracking after the onset of smooth and continuous target
trajectories. In this paper, smooth pursuit was modelled in terms of
hidden causes that attracted both the target and centre of gaze
simultaneously -- enabling the trajectory of the target to inform
estimates of the hidden cause that, in turn, provide predictions about
oculomotor consequences. While this extension accounted for
experimentally observed tracking improvements --under continuous
trajectories -- it does not account for anticipatory movements that have
to accumulate information over time. This anticipatory behaviour could
only be explained with a deeper hierarchical model that has an explicit
representation of latent (periodic) structure causing target motion.
When the generative model was equipped with a deeper structure, it was
then able to produce anticipatory movements of the sort seen
experimentally. Clearly, the simulations in this paper are just
heuristic and do not represent a proper simulation of neurobiological
processing. However, they can be taken as proof of principle that the
basic computational architecture -- in terms of generalized
representations and hierarchical models -- can explain some important
and empirical facts about eye movements. In what follows, we consider
the models in this paper in relation to other models and how modelling
of this sort may have important implications for understanding the
visual-oculomotor system.

\subsection{Comparison with other models}

The model that we have presented here speaks to and complements several
existing models of the oculomotor system. First, it shares some
properties with computer vision algorithms used for image stabilisation.
Such models often use motion detection coupled with salient feature
detection for the registration of successive frames~\citep{Lucas81}. A major difference is that these models are often applied to very
specific problems or configurations for which they give an efficient,
yet \emph{ad hoc} solution. A more generic approach is to use - as our
model does - a probabilistic method, for instance particle filtering~\citep{Isard98}. Our model provides a constructive extension --
as we integrate the dynamics of both sensation and action. In principle,
this could improve the on-line response of feature tracking algorithms.

Second, using our modelling approach, we reproduce similar behaviours
shown by other neuromimetic models of the oculomotor system. For
example, the pursuit of a dot with known uncertainty can be modelled as
the response of a Kalman filter~\citep{Kalman60}. Both generalized Bayesian
(active inference) and Kalman filtering predict the current state of the
system using prior knowledge (about previous target locations) and
refine these predictions using sensory data (prediction errors). This
analogy with block-diagrams from control theory was first highlighted by~\citep{Robinson86} and~\citep{Krauzlis89} -- and has
since been used widely~\citep{Grossberg97}. For a recent treatment involving the neuromorphic modelling of cortical areas, see~\citep{Shibata05}. However, it should be
noted that the link with Kalman filtering is rarely explicit (but see~\citep{deXivry2013Kalman}); most
models have been derived heuristically, rather than as optimal solutions
under a generative model. One class of such neuromimetic models uses
neural networks that mimic the behaviour of the Kalman filter~\citep{Haykin01}. This model was used to fit and predict the response of smooth
pursuit eye movements under different experimental parameters~\citep{Montagnini07} or while interrupting information flow~\citep{Bogadhi11}. Developing this methodology -- and by analogy with
modular control theory architectures -- these building blocks can be
assembled to accommodate increasingly complex behavioural tasks. This
can take the form of a multi-layered model for transparency processing~\citep{Raudies11} or of an interconnected graph connecting the form
and motion pathways~\citep{Beck08b}. Such models have been used to
understand adaptation to blanking periods and to tune the balance
between sensory and proprioceptive inputs~\citep{Madelain03}.
Our model is different in a key aspect: The Kalman filter is indeed the
(Bayes) optimal solution under a linear generative model but a cascade
of such solutions is not the optimal solution to (non-linear)
hierarchical models~\citep{Balaji11}. 
The active inference approach considers the (embodied) system as a whole and furnishes an optimal solution in the form of generalized Bayesian filtering. In particular, given the delays at the sensory and motor levels, it provides an optimal solution that accommodates (or compensates for) these delays. As shown in the results, the ensuing behavior reproduces experimental results from pursuit initiation~\citep{Masson10} to anticipatory responses~\citep{Avila06,Barnes00}. The approach thus provides in inclusive framework, compared with heuristics used in neuromimetic models that focus on specific aspects of oculomotor control (see below). %

The model presented here shares many features with other probabilistic
models. First, representations are encoded as probability density is.
This allows processing and control to be defined in terms of
probabilistic inference; for instance, by specifying a prior belief that
favours slow speeds~\citep{Weiss02}. This approach has been
successful in explaining a wide variety of physiological and
psychophysical results. For example, it allows one to model spatial~\citep{Perrinet07neurocomp} or temporal~\citep{Montagnini07}
integration of information, using conditional independence assumptions.
Furthermore, recent developments have addressed the estimation of the
shape and parameters of priors for slow speeds~\citep{Stocker06} and for the integration of ambiguous versus non-ambiguous
information~\citep{Bogadhi11a}. The active inference scheme used
here relies on generative models that entail exactly the same sorts of
priors. It has also been shown that free energy minimisation extends the
type of probabilistic models described above to encompass retinal
stabilisation and oculomotor reflexes~\citep{Friston10a}. A crucial
difference here is that we have explicitly considered the problem of
dynamics and delays. Our goal was to understand how the system could
provide an optimal solution, when it knows (or can infer) the delay
between sensing input (in the past) and processing information that
informs action (in the future). This endeavour allowed us to build a
model -- using simple priors over the dynamics of the hidden causes --
that reproduces the sorts of anticipatory behaviour seen empirically.

\subsection{Limitations}

Clearly, there are many aspects of oculomotor control we have ignored in
this theoretical work. Foremost, we have used a limited set of stimuli
to validate the model. Pursuit initiation was only simulated using a
simple sweep of a dot, while smooth pursuit was studied using a
sinusoidal trajectory. However, these types of stimuli are commonly used
in the literature, as they best characterise the type of behaviour
(following, pursuit) that we have tried to characterise: see~\citep{Barnes08} for a review. We have not attempted to reproduce the oscillations
at steady state as in~\citep{Robinson86} or~\citep{Goldreich92},
although this may help to optimise the parameters of our model in
relation to empirical data. The hemi-sinusoidal stimulus is also a
typical stimulus for studying anticipatory responses~\citep{Avila06,Barnes00}. Further validations of this model would call on a
wider range of stimuli and consider and accumulated wealth of
neurophysiological and behavioural data~\citep{Tlapale10benchmark}.

In this paper, we have focused on inference under a series of generative models of oculomotor control. We have not considered how these models are acquired or learned. In brief, the acquisition of generative models and their subsequent optimisation in terms of their parameters (i.e., synaptic connection strengths) is an important, if distinct, issue. In the context of active inference, model acquisition and perceptual learning can be cast in terms of model selection and parameter optimisation through the minimisation of free energy. Under certain simplifying assumptions, this learning reduces to associative plasticity. A discussion of these and related issues can be found in~\citet{Friston08b}.

The generative model used in this paper has no explicit representation of space but only the uncertain, vectorial position of a target. We have previously studied the role of prediction in solving problems that are associated with the detection of motion using a dynamical and probabilistic model of spatial integration~\citep{Perrinet12}. Both that model and the current model entertain a similar problem: that of the integration of local information into a global percept, in both the temporal (this manuscript) and spatial~\citep{Perrinet12} domains. We have considered integrating sensory information in the spatial domain: terms of the prediction of sensory causes and their sampling by saccades~\citep{Friston12a}, and of the effects on smooth pursuit of reducing the precision. This manipulation can account for several abnormalities of smooth pursuit eye movements typical of schizophrenia~\citep{Adams12}. In this paper, we have limited ourselves to integrating information over time. It would be nice, in the future, to consider temporal and spatial integration simultaneously.

A final limitation of our model is the simplified modelling of the
physical properties of the oculomotor system -- due to the biophysics of
the eyes and photoreceptors, sensory input contains motion streaks that
can influence the detection of motion~\citep{Barlow04}.
Furthermore, we have ignored delays in neuronal message passing among
and within different levels of the hierarchy: for a review of
quantitative data from monkeys see~\citep{Salin95}. %
Finally, we have not considered in any depth the finer details of how predictive coding or Bayesian filtering might be implemented neuronally. It should be noted that predictive coding in the cortex was attended by some early controversies; for example, paradoxical increases in visual evoked responses were observed when prediction error should be minimal. For example, a match between sensory signals and descending predictions can lead to the enhancement of neuronal firing~\citep{Roelfsema98}. The neuronal implementation assumed in our work (see~\ref{sec:active-inference}) finesses many of these issues. In this (hypothetical) scheme, predictions and prediction errors are encoded by the neuronal activity of deep and superficial pyramidal cells respectively (Mumford 1992; Bastos et al. 2012). In this scheme, the enhancement of evoked responses is generally thought to reflect attentional gain, which corresponds to the optimization of the expected precision (inverse variance) of prediction errors, via synaptic gain control ~\citep{Feldman10a}. Put simply, attention increases the gain of salient or precise prediction errors that the predictions are trying to suppress. Indeed, the orthogonal effects of expectations and attention in predictive coding have been established empirically using fMRI~\citep{Kok12}. See Bastos et al. (2012) for a review of the anatomical and electrophysiological evidence that is consistent with the scheme used here. %
\subsection{Perspectives}
Notwithstanding the limitations above, this approach may provide some interesting perspectives on neural computations in the oculomotor system. First, the model presented here can be compared to existing models of the oculomotor system. In particular, any commonalities of function suggest that extant neuromimetic models may be plausibly implemented using a generic predictive coding architecture.  Second, the Bayes optimal control solution rests on a computational (anatomical) architecture that can be informed by electrophysiological or psychophysical studies. For example, we have considered only delays at the motor and sensory level. However, delays in axonal conduction between hierarchical levels -- within the visual-oculomotor system -- may have implications for intrinsic and extrinsic connectivity: in visual search, predictions generated in higher areas (say supplementary and frontal eye fields) may exploit a shorter path, by stimulating the actuator to sample more information (by making an eye movement) rather than accumulating evidence by explaining away prediction errors in lower (striate and extrastriate) cortical levels~\citep{Masson10}. By studying the structure of connections implied by theoretical considerations (see Figure~\ref{fig:figure2}), our modelling approach could provide a formal framework to test these sorts of hypotheses. A complementary approach would be to apply dynamic causal modelling~\citep{Friston03a} to electrophysiological data, using predictive coding architectures, such that transmission delays (and their compensation or modeling) among levels of the visual-oculomotor system could be evaluated empirically. %
A recent example of using dynamic causal modelling to test hypotheses based upon predictive coding architectures can be found in Brown and Friston (2013). This example focuses on attentional gain control in visual hierarchies.

Second, this work may provide a new perspective for experiments, in
particular for the generation of stimuli. We have previously considered
such a line of research by designing naturalistic, texture-like
pseudo--random visual stimuli to characterise spatial integration during
visual motion detection~\citep{Leon12a}. We were able to show that
the oculomotor system exhibits an increased following gain, when stimuli
have a broad spatial frequency bandwidth. Interestingly, the velocities
of these stimuli were harder to discriminate relative to narrow
bandwidth stimuli -- in a two alternative forced-choice psychophysical
task~\citep{Simoncini12}. In this work, the authors used competitive
dynamics based on divisive normalisation. Moreover, textured stimuli
were based on a simple forward model of motion detection~\citep{Leon12a}. This may call for the use of more complex generative models to
generate such textures. In addition, the use of gaze contingent
eye-tracking systems allows real-time manipulation of the configuration
(position, velocity, delays) of the stimulus, with respect to eye
position and motion. By targeting different sources of uncertainty, at
the different levels of the hierarchical model, one might be able to get
a better characterisation of the oculomotor system.

The confounding influence of delays inherent in neuronal processing is a
strong biophysical constraint on neuronal dynamics. Representations in
generalized coordinates of motion provide a potential resolution that
may have enjoyed positive evolutionary pressure. However, it remains
unclear how neural information, represented in a distributed manner
across the nervous system, is integrated with exteroceptive, operational
time. The ``binding'' of different information, without a central clock,
seems essential, but the correlate of such a temporal representation of
sensory information (independent of delays) has never been observed
explicitly in the nervous system. Elucidating the neural representation
of temporal information would greatly enhance our understanding of both
neural computations themselves and our interpretation of measured
electromagnetic (EEG and MEG) signals that are tightly coupled to those
computations.

\section*{Acknowledgments}
\Acknowledgments
\section{Appendix}
\subsection{Appendix 1: Variational free energy}

Here, we derive various formations of free-energy and show they relate to each other. We start with the quantity we want to bound and implicitly minimise --- namely, surprise or the negative log-evidence associated with sensory states $\tilde{s}(t)$ that have been caused by some unknown quantities $\Psi(t)$. These hidden causes correspond to the (generalized) motion (that is, position, velocity, acceleration, ...) of a target that the oculomotor system is tracking.

\begin{align}
- \ln p(\tilde{s}) = - \ln  \int p(\tilde{s} , \Psi) d \Psi \label{eq:app1-1}
\end{align}%

We now simply add a non-negative cross-entropy or divergence between some arbitrary (conditional) density $q(\Psi)=q(\Psi | \tilde{\mu})$ and the posterior density $p(\Psi | \tilde{s})$ to create a free-energy bound on surprise 

\begin{align}
F &= - \ln p(\tilde{s}) + \int  q(\Psi) \ln \frac{ q(\Psi)}{ p(\Psi | \tilde{s})} d \Psi \nonumber \\%
 &= - \ln p(\tilde{s}) + D( q(\Psi) || p(\Psi | \tilde{s}) )   \label{eq:app1-2}
\end{align}%

The cross entropy term is non-negative by Gibb's inequality. Because surprise depends only on sensory states, we can bring it inside the integral and use $p(\tilde{s} , \Psi) = p(\Psi | \tilde{s}) p(\tilde{s})$ to show free-energy is a Gibb's energy $G=- \ln p(\tilde{s} , \Psi)$ expected under the conditional density minus the entropy of the conditional density

\begin{align}
F &= \int  q(\Psi) \ln \frac{ q(\Psi)}{  p(\Psi | \tilde{s}) p(\tilde{s})} d \Psi \nonumber \\%
 &= \int  q(\Psi) \ln \frac{ q(\Psi)}{  p(\Psi, \tilde{s})} d \Psi  \label{eq:app1-3} \\%
 &= -\int  q(\Psi) \ln p(\Psi, \tilde{s}) d \Psi + \int q(\Psi) \ln  q(\Psi) d \Psi \nonumber %
\end{align}%

This is a useful formulation because it can be evaluated in a relatively straightforward way given a probabilistic generative model $p(\tilde{s} , \Psi)$. A final rearrangement, using $p(\tilde{s} , \Psi) = p(\tilde{s} | \Psi) p(\Psi)$, shows free-energy is also complexity minus accuracy, where complexity is the divergence between the recognition density $q(\Psi)$ and the prior density $p(\Psi)$

\begin{align}
F &= \int  q(\Psi) \ln \frac{ q(\Psi)}{  p(\Psi | \tilde{s}) p(\tilde{s})} d \Psi \nonumber \\%
 &= -\int  q(\Psi) \ln p(\tilde{s} | \Psi) d \Psi + D( q(\Psi) || p(\Psi) ) \label{eq:app1-4} %
\end{align}%

\subsection{Appendix 2: The maximum entropy principle and the Laplace assumption}

If we admit an encoding of the conditional density up to second order moments, then the maximum entropy principle~\citep{Jaynes57}, implicit in the definition of free energy above, requires $q(\Psi | \tilde{\mu}) = \mathcal{N}(\tilde{\mu}, \Sigma)$ to be Gaussian. This is because a Gaussian density has the maximum entropy of all forms that can be specified with two moments. Assuming a Gaussian form is known as the Laplace assumption and enables us to express the entropy of the conditional density in terms of its first moment or expectation. This follows because we can minimise free energy with respect to the conditional covariance as follows:

\begin{align} 
F &= G(\tilde{s}, \tilde{\mu}) + \frac{1}{2} tr( \Sigma \partial_{\tilde{\mu} \tilde{\mu}} G ) - \frac{1}{2} \ln | \Sigma |  \nonumber \\%
G &= - \ln p(\tilde{s} , \Psi) \nonumber \\%
\partial_\Sigma F &= \frac{1}{2} \partial_{\tilde{\mu} \tilde{\mu}} G  - \frac{1}{2} \Pi 
\label{eq:app2-1} %
\end{align}%
so that $\partial_\Sigma F = 0$ implies
\begin{align}
\Pi &= \partial_{\tilde{\mu} \tilde{\mu}} G \nonumber \\%
F &= G(\tilde{s}, \tilde{\mu}) + \frac{1}{2} \ln | \partial_{\tilde{\mu} \tilde{\mu}} G | \label{eq:app2-2} %
\end{align}%

Here, the conditional precision $\Pi(\tilde{s}, \tilde{\mu})$ is the inverse of the conditional covariance $\Sigma(\tilde{s}, \tilde{\mu})$. In short, free energy is a function of generalized conditional expectations and sensory states.

\subsection{Appendix 3: Integrating or solving active inference schemes using generalized descents.}

Given a generative model or its associated Gibbs energy function, one can now simulate active inference by solving the following set of ordinary differential equations for a system that includes generalized real-world states and internal states of the agent mediating (delayed) action and perception:

\begin{align}
\dot{u}  = %
\left[\begin{array}{c} 
\dot{\tilde{\bm{s}}} \\ %
\dot{\tilde{\bm{x}}} \\ %
\dot{\tilde{\bm{\nu}}} \\  %
\dot{\bm{\tilde{\omega}_\nu}} \\  %
\dot{\bm{\tilde{\omega}_x}} \\  %
\dot{\tilde{\mu}} \\  %
\dot{\tilde{\eta}}\\  %
\dot{\bm{a}}  %
\end{array}\right] = %
\left[\begin{array}{c} 
\mathcal{D}\tilde{\bm{g}}(\tilde{\bm{x}}, \tilde{\bm{\nu}}, \tilde{\bm{a}}) + \mathcal{D}\bm{\tilde{\omega}_\nu} \\%
\tilde{\bm{f}}(\tilde{\bm{x}}, \tilde{\bm{\nu}}, \tilde{\bm{a}}) + \bm{\tilde{\omega}_x} \\%
\mathcal{D}\tilde{\bm{\nu}} \\%
\mathcal{D}\bm{\tilde{\omega}_\nu} \\%
\mathcal{D}\bm{\tilde{\omega}_x} \\%
\mathcal{D}\tilde{\mu} - \partial_{\tilde{\mu}} F( T(\tau_s -\bm{\tau_s}) \bm{\tilde{s}}, \tilde{\mu})\\  %
\mathcal{D}\tilde{\eta}\\  %
- \partial_{a} F( T(\tau_s -\bm{\tau_s} + \tau_a -\bm{\tau_a}) \bm{\tilde{s}}, T(\tau_a -\bm{\tau_a}) \tilde{\mu})  %
\end{array}\right]  \label{eq:app3-1} %
\end{align}%

generalized action $\bm{\tilde{a}}(t)$ is approximated using discrete values of $\bm{a}(t)$  from the past. Note that we have included a prior expectation $\tilde{\eta}(t)$ of hidden causes to complete the agent's generative model of its world. Integrating or solving Equation~\ref{eq:app3-1} corresponds to simulating active inference. The updates of the collective states over time steps of $\Delta t$ use a local linearisation scheme~\citep{Ozaki92}:

\begin{align}
\Delta u &=(\exp (\Delta t \cdot \partial_u \dot{u}) - I)(\partial_u \dot{u})^{-1} \nonumber \\%
\partial_u \dot{u} &= \nonumber \\%
& \left[\begin{array}{cccccccc}
0 & \mathcal{D}\partial_{\tilde{\bm{x}}} \tilde{\bm{g}}  & \mathcal{D}\partial_{\tilde{\bm{\nu}}} \tilde{\bm{g}} & \mathcal{D} & \ldots & & & \mathcal{D}\partial_a\tilde{\bm{g}}  \\%
 & \partial_{\tilde{\bm{x}}} \tilde{\bm{f}}  & \partial_{\tilde{\bm{\nu}}} \tilde{\bm{f}} &  & I  & & & \partial_a\tilde{\bm{f}}  \\%
\vdots &  & \mathcal{D} & & \vdots & \vdots & &  \\%
 & &  & \mathcal{D}&  &  &  & \\%
 & & \ldots & & \mathcal{D} &  & \ldots & \\%
- \partial_{\tilde{\mu}\tilde{s}} F &  & \ldots &  & & -\mathcal{D}\partial_{\tilde{\mu}\tilde{\mu}} F & - \partial_{\tilde{\mu}\tilde{\eta}} F & -\partial_{\tilde{\mu}a} F  \\%
 &  & &  & & -\partial_{\tilde{\eta}\tilde{\mu}} F & \mathcal{D} &   \\%
- \partial_{a\tilde{s}} F &  & &  & & - \partial_{a\tilde{\mu}} F & - \partial_{a\tilde{\eta}} F &  - \partial_{aa} F %
\end{array}\right]  \label{eq:app3-2}%
\end{align}%

Details about how to compute the gradients and curvatures pertaining to the conditional expectations can be found in~\citep{Friston10d}. These are generally cast in terms of prediction errors using straightforward linear algebra. Because action can only affect free-energy through the sensory states, its dynamics are prescribed by the following gradients and curvatures (ignoring higher-order terms): 

\begin{align}
\partial_a F &= (\partial_a \tilde{\varepsilon}^{(1)}_\nu)  \cdot  \Pi^{(1)}_a T(\tau_a -\bm{\tau}_a)  \tilde{\varepsilon}^{(1)}_\nu \nonumber  \\%
\partial_{aa} F &= (\partial_a \tilde{\varepsilon}^{(1)}_\nu)  \cdot  \Pi^{(1)}_a T(\tau_a -\bm{\tau}_a)  (\partial_a \tilde{\varepsilon}^{(1)}_\nu) \nonumber  \\%
\label{eq:app3-3} \\%
\partial_a \tilde{\varepsilon}^{(1)}_\nu &= T(\tau_s-\bm{\tau}_s)  \partial_a \bm{\tilde{s}}(t) \nonumber \\%
\partial_a  \bm{\tilde{s}} &= \partial_a  \bm{\tilde{g}} + \partial_{\tilde{\bm{x}}} \bm{\tilde{g}}( \sum_i \mathcal{D}^{-i}(\partial_{\tilde{x}} \tilde{\bm{f}}) ^{i-1}) \partial_{a} \tilde{\bm{f}} \nonumber  %
\end{align}%

The partial derivative of the sensory states with respect to action and is specified by the generative process. In biologically plausible instances of this scheme, this derivative would have to be computed on the basis of a mapping from action to sensory consequences. It is generally assumed that agents are equipped with $\partial_a  \bm{\tilde{s}} $ epigenetically, because it has a simple form. For example, contracting a muscle fibre elicits a proprioceptive stretch signal in a one-to-one fashion. The precision matrix $ \Pi^{(1)}_a$ in Equation~\ref{eq:app3-3} is specified such that only proprioceptive prediction errors with these simple forms have nonzero precision. This can be regarded as the motor gain in response to proprioceptive prediction errors. 
Equation~\ref
{eq:app3-2} may look complicated but can be evaluated automatically using numerical derivatives for any given generative model. All the simulations in this paper used just one routine ---$\mathsf{toolbox/DEM/spm\_ADEM.m}$---  and summarised in the script  $\mathsf{toolbox/DEM/ADEM\_oculomotor\_delays.m}$. Both are available as part of the SPM software (\url{http://www.fil.ion.ion.ucl.ac.uk/spm}).%
\printbibliography%

\end{document}